\pdfoutput=1
\documentclass[a4paper,american,aps,citeautoscript,floatfix,pdftex,prl,showpacs,superscriptaddress,%
reprint,twocolumn]{revtex4-1}
\usepackage{amsfonts,amsmath,amssymb}
\usepackage{graphicx}
\usepackage[T1]{fontenc}
\usepackage[utf8]{inputenc}
\usepackage{microtype}
\usepackage{pxfonts,tgpagella}
\usepackage{soul}
\usepackage{subfigure}
\usepackage{textcomp}
\usepackage[textsize=footnotesize]{todonotes}
\usepackage{xspace}
\usepackage{hyperref, hypernat}
\usepackage[todos]{CMImacros}

\newlength{\figwidth}
\newcommand{\cfeldesy}{\affiliation{Center for Free-Electron Laser Science, DESY, Notkestrasse 85,
      22607 Hamburg, Germany}}%
\newcommand{\uhhcui}{\affiliation{The Hamburg Center for Ultrafast Imaging, University of Hamburg,
      Luruper Chaussee 149, 22761 Hamburg, Germany}}%
\newcommand{\uhhphys}{\affiliation{Department of Physics, University of Hamburg, Luruper Chaussee
      149, 22761 Hamburg, Germany}}%

\begin{document}
\title{Electron gun for diffraction experiments off controlled molecules}%
\author{Nele L.\,M.\ Müller}\cfeldesy%
\author{Sebastian Trippel}\cfeldesy%
\author{Karol Długołęcki}\cfeldesy%
\author{\mbox{Jochen~Küpper}}%
\email{jochen.kuepper@cfel.de}%
\homepage{https://www.controlled-molecule-imaging.org}%
\cfeldesy\uhhcui\uhhphys%
\date{\today}%
\begin{abstract}\noindent%
   A dc electron gun, generating picosecond pulses with up to $8\times10^{6}$ electrons per pulse,
   was developed. Its applicability for future time-resolved-diffraction experiments on state- and
   conformer-selected laser-aligned or oriented gaseous samples was characterized. The focusing
   electrodes were arranged in a velocity-map imaging spectrometer configuration. This allowed to
   directly measure the spatial and velocity distributions of the electron pulses emitted from the
   cathode. The coherence length and pulse duration of the electron beam were characterized by these
   measurements combined with electron trajectory simulations. Electron diffraction data off a thin
   aluminum foil illustrated the coherence and resolution of the electron-gun setup.
\end{abstract}
\maketitle%
\noindent%
Diffractive imaging is a promising approach to unravel the microscopic details of chemical processes
through the recording of so-called molecular movies in the gas-phase, which trace the structural
dynamics of individual molecules and nano-particles at the atomic level. Electron and x-ray
diffraction are well established tools to investigate the structures of solid state
samples~\cite{Germer:PR56:58}, for example in transmission electron
microscopy~\cite{Takayanagi:JVS3:1502} or x-ray crystallography~\cite{Lonsdale:PRSA123:494,
   Yonath:ARB1:257}. Furthermore, electron diffraction has found broad application for gas-phase
structure-determination in chemistry~\cite{Hargittai:GED:1988}. Recent developments have mainly
focused on realizing time-resolved experiments in order to study structural dynamics, where x-ray
and electron diffraction served as complementary approaches~\cite{Williamson:Nature386:159,
   Ihee:Science291:458, Siwick:Science302:1382, Chapman:NatPhys2:839, Spence:RPP75:102601,
   Barty:ARPC64:415}.

To be able to record structural changes during ultrafast molecular processes of small complex
molecules in the gas phase, signals from many identical molecules have to be averaged. Gas-phase
investigations pose the challenge that the sample might comprise different isomers and
sizes~\cite{Chang:IRPC:inprep}. In addition, the molecules in the gas phase are typically randomly
oriented. It is therefore important to provide samples, which are as clean and defined as possible,
to allow for experimental averaging over multiple electron pulses. Clean molecular samples can be
generated by their spatial separation according to shape~\cite{Filsinger:ACIE48:6900,
   Kierspel:CPL591:130, Filsinger:PRL100:133003} and size~\cite{Trippel:PRA86:033202}. Controlling
the spatial orientation of the molecules leads to an enhancement of the information that can be
retrieved from a diffraction pattern, as proposed theoretically~\cite{Ryu:JPCA107:6622,
   Spence:PRL92:198102, Pabst:PRA81:043425, Filsinger:PCCP13:2076} and demonstrated experimentally
for x-ray~\cite{Kuepper:PRL112:083002, Stern:FD171:393} as well as for electron
diffraction~\cite{Hensley:PRL109:133202, Yang:SD1:044101}. Strong alignment or orientation is
generally necessary~\cite{Filsinger:PCCP13:2076, Barty:ARPC64:415} for three-dimensional structure
reconstruction~\cite{Yang:SD1:044101} and can be provided in cold supersonic molecular beams by
strong-field laser alignment and mixed-field orientation~\cite{Stapelfeldt:RMP75:543,
   Holmegaard:PRL102:023001, Ghafur:NatPhys5:289, Kraus:PRL113:023001, Trippel:PRL114:103003}. The
low density of these controlled gas-phase samples requires sources of large-cross-section particles
or photons with large brightness, while still ensuring atomic resolution. Electron sources can meet
these requirements even in table-top setups.

The first sources for creating electron pulses short enough to study ultrafast processes in
molecules or materials were dc electron guns. Here, electrons were created from metallic surfaces by
short laser pulses and accelerated in dc electric fields~\cite{Ihee:Science291:458,
   Siwick:Science302:1382}, yielding sub-picosecond electron pulses of moderate coherence and
brilliance. Radio-frequency cavities allow for temporal compression of electron pulses through
phase-space rotation, shortening the pulse duration to below 100~fs with electron numbers of
$10^{6}$ per pulse and electron spot sizes below 100~\um~\cite{vanOudheusden:PRL105:264801}. Compact
dc guns can achieve comparable properties by increasing the acceleration fields and reducing the
path length, during which the electron pulse can expand~\cite{Sciaini:RPP74:096101,
   Musumeci:RSI:013306, Robinson:RSI86:013109, Gerbig:NJP17:043050}. Ultra-fast-single-electron
sources~\cite{Lahme:SD1:034303} avoid the problem of space charges, but rely on very high repetition
rates to achieve sufficient electron fluxes for diffractive imaging experiments. The use of ultra
cold atoms as electron sources increases the coherence~\cite{vanMourik:SD1:034302,
   McCulloch:NatPhys10:785}. Other possible sources for time-resolved electron diffraction are low
energy electron setups~\cite{Gulde:Science345:200} or laser-induced electron
diffraction~\cite{Zuo:CPL159:313, Blaga:Nature483:194}.

If the molecular samples are prepared in the necessary strongly-controlled fashion, their densities
are typically on the order of some $10^8~\text{molecules/cm}^{3}$~\cite{Kuepper:PRL112:083002,
   Chang:Science342:98}. Assuming Rutherford scattering, for the prototypical molecule
2,5-diiodobenzonitrile an effective cross section on the order of $10^{-15}~\text{cm}^2$ can be
derived for our experimental geometry and a beam stop blocking a solid angle of
$1.3\times10^{-3}$~sr. To align or orient the molecules, they are typically exposed to laser fields
with intensities of $1~\text{TW}/\text{cm}^2$~\cite{Stapelfeldt:RMP75:543}, which can be achieved by
focusing the ps-duration mJ-pulse-energy laser beam to 100~\um~\cite{Trippel:MP111:1738}. For a
500~\um thick molecular beam this results in an interaction volume of about
$5\times10^{-6}~\text{cm}^{3}$. The number of molecules in this volume and the given cross section
lead to an elastic scattering signal on the detector $S^D_\text{elastic}$ of $5\times10^{-9}$ per
electron. In order to achieve a diffraction pattern containing some $10^5$ scattered electrons on
the detector within 24 hours, bright electron sources with $10^{9}$ electrons per second are needed.
For experimental repetition rates on the order of 1~kHz~\cite{Trippel:MP111:1738}, this corresponds
to $10^{6}$ electrons per shot with an electron beam focus size of approximately 100~\um. The setup
presented here produced the necessary electron numbers and allowed for a characterization of the
electron beam to ensure, for instance, the required transverse coherence length.

\begin{figure}
   \centering
   \includegraphics[width=\linewidth]{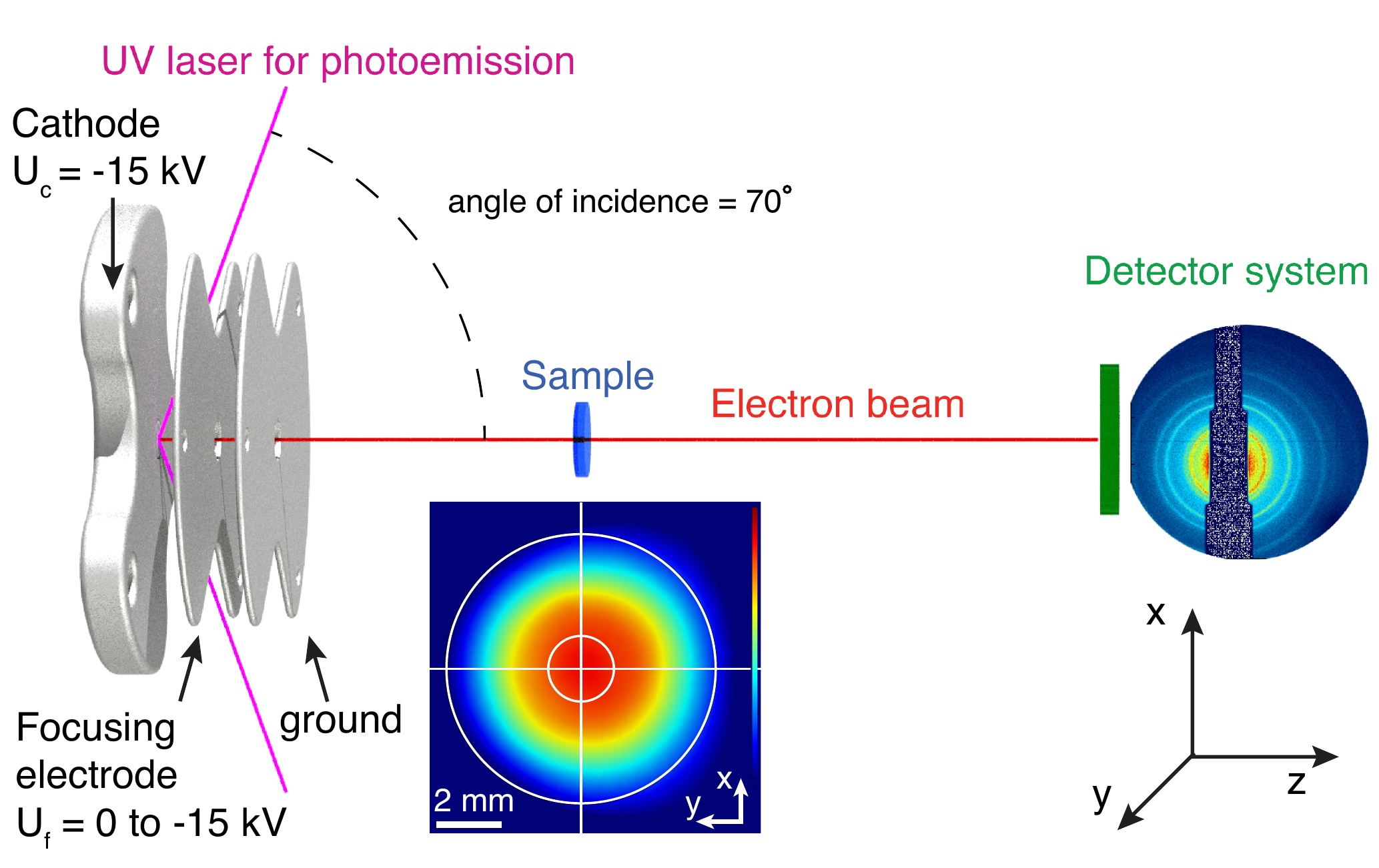}
   \caption{Experimental scheme of the electron gun consisting of three electrodes in
      velocity-map-imaging-spectrometer configuration. After photoemission, the electrons were
      diffracted of an aluminum sample or were measured by a Faraday cup coupled to an electrometer.
      A multi-channel-plate detector with phosphor screen and a camera was used for position
      sensitive detection. The inset at the bottom depicts the potential between cathode and
      focusing electrode. To highlight the asymmetry the $x=0$ and $y=0$ axes as well as centered
      circles are shown to guide the eye.}
   \label{fig:setup}
\end{figure}
A schematic of the experimental setup is shown in \autoref{fig:setup}. The electrons were
photo-emitted from a copper cathode \emph{via} one-photon absorption after irradiation with short UV
laser pulses. The pulses were generated by third-harmonic generation (THG) of 30-fs-duration
near-infrared pulses from a Ti:Sapphire laser (TSL) system with a repetition rate of 1~kHz. Based on
the pulse duration of the near-infrared laser pulse at the THG setup, and the dispersion in
subsequent optical elements, we estimated a pulse duration of 370~fs for the UV light. The central
wavelength was 265~nm with a spectral width of 4~nm. The light enters and exits the chamber through
anti-reflection coated windows. The pulse impinged on the cathode under an angle of \degree{70} to
the surface normal of the cathode. The electrons were accelerated and focused by three electrodes in
velocity-map imaging spectrometer (VMI) configuration. The applied potential at the cathode was
$U_c=-15$~kV. The voltage on the focusing electrode was varied between $U_f=0~\text{kV}$ and
$-15~\text{kV}$. The third electrode was kept on ground potential. The asymmetric electrode shape
allowed the laser beam to pass and impinge on the cathode. The holes within the electrodes were
large enough to avoid clipping of the electron beam. On the one hand this reduced the background
signal in diffraction experiments, as there was no electron scattering off the electrodes. On the
other hand it allowed for steering the electron beam's position by changing the laser-spot position
on the cathode. The voltage on the focusing electrode $U_f$ determined the position of the electron
beam focus along the $z$ direction. The electrode configuration allowed to characterize the electron
pulse by applying the corresponding voltages for spatial- and
velocity-mapping~\cite{Eppink:RSI68:3477, Stei:JCP138:214201, Bainbridge:NJP16:103031}. It is
possible to create electric fields that allow to either map the spatial distribution of the
electrons at the cathode or their respective velocity distribution as a 2D projection onto the
detector~\cite{Eppink:RSI68:3477, Stei:JCP138:214201}. The spatial distribution of the electron beam
was recorded by a position sensitive detector consisting of a multi-channel-plate (MCP) with a
phosphor screen and a CMOS camera (Optronis CamRecord CL600x2). The detection system was read out
with a 1~kHz repetition rate, which allowed for single electron counting in the case of a few
electrons per pulse. At large electron numbers, the gain of the detector had to be reduced to avoid
damage of the MCP. With lower gain single electrons could not be resolved anymore. In order to
reduce background from scattered light or other sources the detector can be gated. A Faraday cup
connected to an electrometer (Keithley 6514 electrometer) was used to measure the electron number
per pulse. To further characterize the electron pulses, we performed diffraction experiments with a
thin aluminum foil on a TEM grid, which was introduced into the electron beam path. In this case the
direct electron beam was blocked by a copper or aluminum beam block.

The electron gun was designed for ultra-high vacuum. Here, the final pressure was
$4\times10^{-9}$~mbar using a turbomolecular pump with pumping speed of 300~l/s, limited by
outgassing from the cable of the Faraday cup and from PEEK material of the electron-gun insulators.
This low pressure is essential when investigating thin gas-phase samples in order to reduce the
scattering by background gas. We expect to achieve pressures of a few $10^{-10}$~mbar in the final
setup when replacing all PEEK insulators by MACOR or alumina.

\begin{figure}
  \centering
  \includegraphics[width=\linewidth]{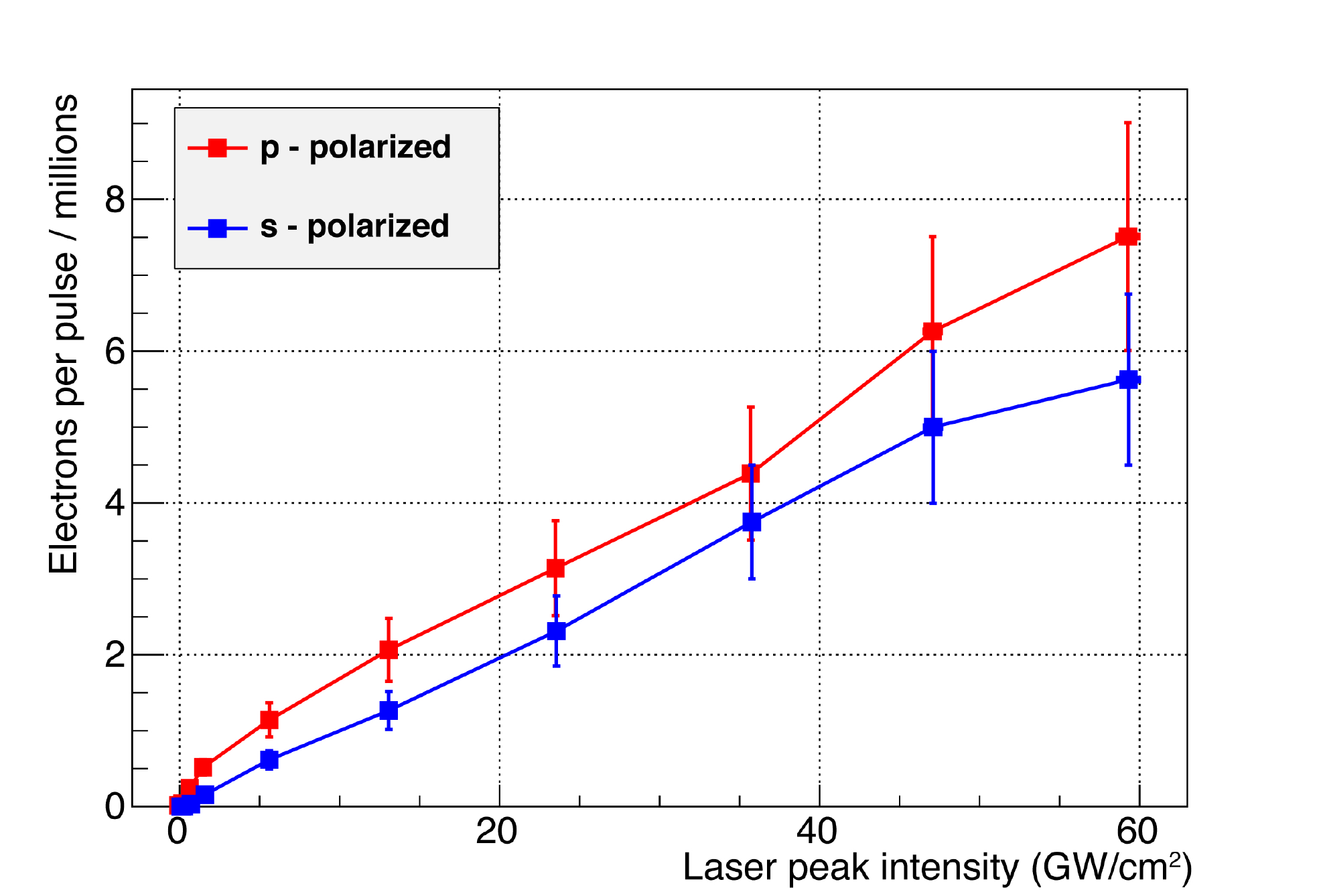}
  \caption{Electron number as a function of laser peak intensity for p-polarized (red) and
     s-polarized (blue) laser light and a central laser wavelength of 265~nm.}
  \label{fig:power}
\end{figure}
In \autoref{fig:power} the electron number per pulse is shown for $U_f=-13.2~\text{kV}$ as a
function of the laser pulse intensity for two laser polarizations. This focusing voltage
corresponded to the focus of the electron beam being close to the detector surface. The number of
electrons increased linearly with the laser power, as expected for a one-photon effect of 265~nm
light with a spectral width of 4~nm on copper, which has a work function of 4.7~eV. No saturation
was observed. The number of generated electrons depended on the laser polarization: For
p-polarization (red curve, field vector in plane of incidence) more electrons were generated than
for s-polarization (blue curve, field vector parallel to cathode surface). This is in accord with
the reflectivity of copper being higher for s-polarized light than for p-polarized light, which was
confirmed by measuring the laser power for both polarizations after the cathode. Up to
$8\times10^{6}$ electrons per shot could be obtained, which is sufficient for the planned
diffraction experiments on dilute gas-phase samples delivered by the controlled-molecules apparatus.

\begin{figure}
   \centering
   \includegraphics[width=\linewidth]{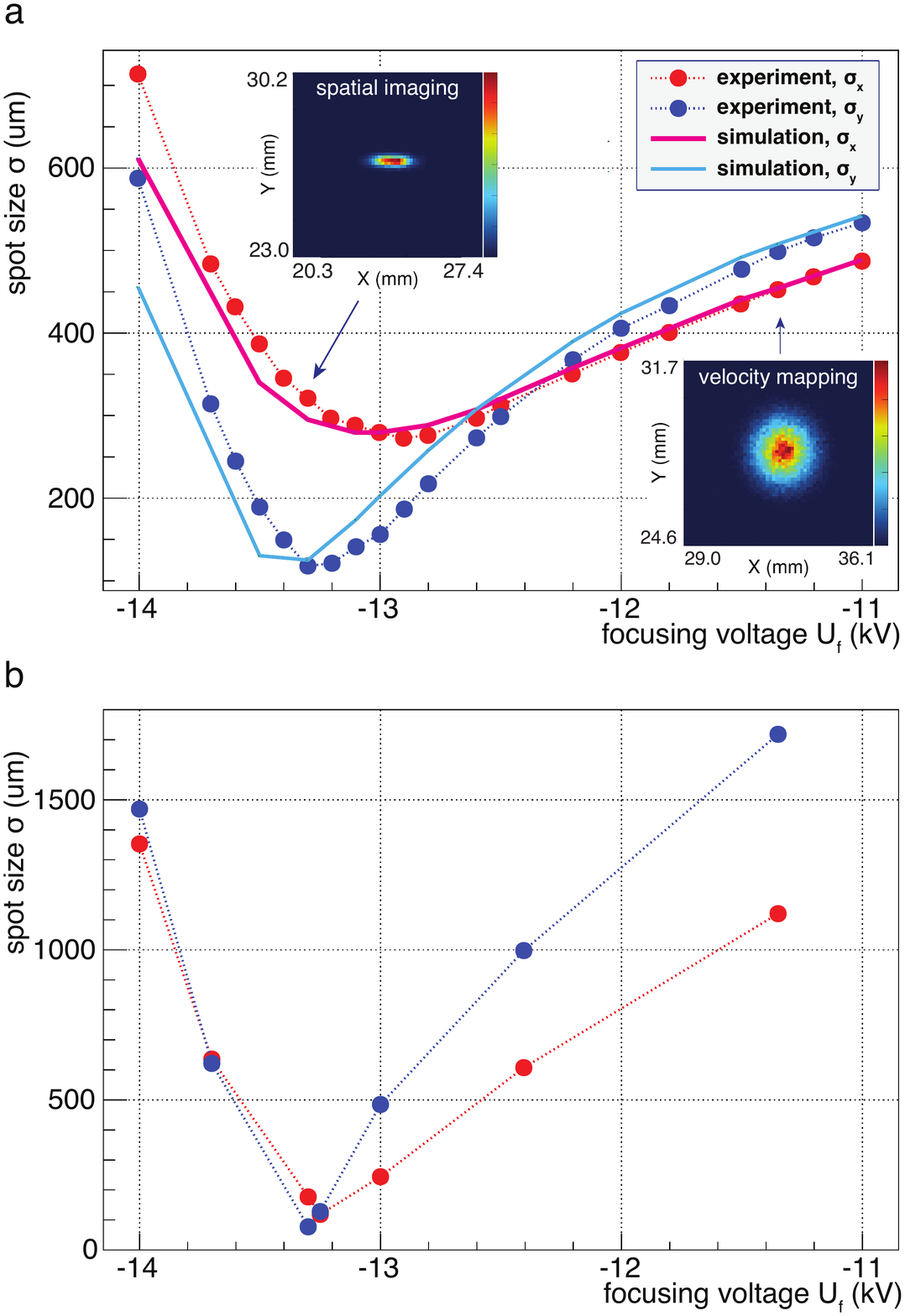}
   \caption{Electron spot size on the detector for different focusing voltages for (a) a few and (b)
      $2\times10^{5}$~electrons per pulse. Red and blue dots are experimental results corresponding
      to $x$ and $y$ directions, respectively. Magenta and cyan lines correspond to corresponding
      simulations in $x$ and $y$ dimension. The insets in (a) show detector images for spatial
      (left) and velocity map imaging (right).}
   \label{fig:VMI}
\end{figure}
For a full characterization of the electron beam the electron spot size at the detector was measured
for various focusing voltages $U_f$, including those for spatial- and velocity-mapping. In
\autoref{fig:VMI}\,a the root-mean-square (RMS) spot size of the electron beam in $x$ and $y$
dimension, $\sigma_x$ and $\sigma_y$, are plotted as a function of $U_f$. Here, the laser intensity
was reduced to less than 10~MW/cm$^2$ to create approximately five electrons per pulse from the
cathode and, therefore, space charge effects were negligible. The spot size decreased with
increasing $U_f$ until it reached a focus, at the detector, for about $U_f=-13$~kV. Raising $U_f$
further led to a defocusing of the electron beam, \ie, the focus was placed before the detector. The
exact voltage to place the focus onto the detector depends on the initial size of the electron
cloud. The electron beam was broadened in $x$ direction due to the large angle of incidence of the
laser. Therefore, the foci in $x$ and $y$ dimension had slightly different focusing behavior.

In spatial imaging mode, $U_f=-13.3$~kV, the spatial distribution of emitted electrons was mapped
onto the detector, which is depicted as inset in \autoref{fig:VMI}\,a. In this case, all electrons,
which started from a certain coordinate on the cathode hit a corresponding point on the detector, in
first order independent of their momentum~\cite{Stei:JCP138:214201}. The magnification factor $m_s$
for spatial imaging was calibrated in the experiment by translating the laser-focus spot on the
cathode using the focusing lens. From a known displacement of the electron beam on the cathode
$\Delta{x}_{C}$ and the corresponding measured displacement of the electrons on the detector
$\Delta{x}_{D}$ a magnification factor of $m_s=\Delta{x}_{D}/\Delta{x}_{C}=3.9$ was determined. This
agreed with the simulated value. For simulations of electric fields and trajectories finite-element
methods were used (COMSOL Multiphysics). The inferred RMS sizes of the electron beam at the cathode
were $\sigma_x=85(3)~\um$ and $\sigma_y=31(1)~\um$; values in parenthesis depict one standard
deviation. The difference in spread originated from the laser impinging on the cathode under an
angle of incidence of \degree{70}. The angle lead to an effective broadening of the photoemission
laser by a factor of approximately three in $x$ direction, while the $y$ dimension was unchanged.
Thus, the created electron beam was broader in $x$ direction than in $y$ direction on the cathode,
which was confirmed in the spatial imaging measurements.

In velocity map imaging mode ($U_f=-11.35$~kV) the transverse velocity distribution was mapped onto
the detector, which is shown as the second inset in \autoref{fig:VMI}\,a. The velocity spread was
similar in both dimensions. With the simulated magnification factor of $m_v=0.9$ for velocity
mapping and a simulated electron time of flight of $4.1$~ns an energy spread of
$\sigma_E=0.1~\text{eV}$ was obtained. This agrees well with the previously reported value
$\sigma_E=0.13~\text{eV}$~\cite{Maldonado:APL101:231103}.

In order to characterize the electron beam further simulations and measurements at various focusing
voltages $U_f$ were performed. The spatial and velocity distribution of the electrons in $x$ and $y$
dimension could be retrieved from the experiment, but the corresponding values in $z$ dimension had
to be simulated. Electric fields were calculated using finite-element methods (COMSOL Multiphysics)
and the electron trajectories in these fields were simulated using
ASTRA~\cite{Floettmann:ASTRA:1997}. The initial spatial distribution at the cathode was taken from
the measurements described above. Together with a Fermi Dirac distribution for the one-photon
emission, this led to the emittance values of
\mbox{$\epsilon_{x}=0.026~\pi\,\text{mrad}\,\text{mm}$} and
\mbox{$\epsilon_{y}=0.010~\pi\,\text{mrad}\,\text{mm}$}, and the energy spread in $z$-direction of
\mbox{$\sigma_{E_z}=0.2~\text{eV}$}. Fitting the emittance to the transverse velocity distributions
retrieved from VMI mode, while keeping $\sigma_{E_z}$ constant, resulted in
\mbox{$\epsilon_{x}=0.029~\pi\,\text{mrad}\,\text{mm}$} and
\mbox{$\epsilon_{y}=0.012~\pi\,\text{mrad}\,\text{mm}$}, in good agreement with the values obtained
from the Fermi Dirac distribution. Using the fitted input parameters, the overall dependence of the
electron beam spot size at the detector on the focusing voltage was simulated. The results are
depicted by the magenta and cyan lines in \autoref{fig:VMI}\,a and are, again, in good agreement
with the experimental results. This indicates that also the simulated $\sigma_{E_z}$ was sensible.
Due to the good agreement between experiment and simulation it is possible to deduce properties of
the electron beam from the simulations, including size, coherence length and pulse duration, for its
whole propagation. The coherence length $L_c=\hbar\,\sigma_x/(m_0\,c\,\epsilon)$, with the electron
mass $m_0$ and the speed of light $c$, was determined using ASTRA~\cite{Floettmann:ASTRA:1997}. At
the sample position (11~cm downstream from the cathode) $L_c$ was deduced to be 3~nm in
$x$-dimension and 1.2~nm in $y$-dimension. The pulse duration at this position was simulated to be
1.4~ps.

\autoref{fig:VMI}\,b shows the spot sizes for $2\times10^5$ electrons per pulse, where space charges
had a significant effect. For the detection of $2\times10^{5}$ electrons per pulse in
\autoref{fig:VMI}\,b, the detector voltage had to be reduced and single electron detection was not
possible. This implies that \autoref{fig:VMI}\,a and b are only qualitatively comparable. A stronger
asymmetry in velocity map imaging mode was observed than above. This could not be reproduced using a
cylindrical symmetry in electric fields and initial velocities, which was a good approximation in
the simulations for few electrons. Using finite-element simulations it was possible to qualitatively
determine the origin of the asymmetry in the velocity map imaging mode, but a full simulation of all
3D trajectories for $2\times10^5$ electrons was not possible due to too high computational cost.
Simulations for few electrons showed that the trajectories of the electrons far off the central axis
of the spectrometer were disturbed by the asymmetry of the electric field due to the opening in the
electrodes, see inset in \autoref{fig:setup}. This became more pronounced when space charges lead to
a significant broadening of the electron distribution. In the case of $2\times10^5$ electrons per
pulse the radial distribution between the cathode and the focusing electrode was increased by an
order of magnitude compared to the few electron case. This lead to a larger magnification factor in
vertical direction in velocity map imaging mode and, therefore, contributed to the asymmetry in the
detector image. Secondly, the space charge effect itself lead to an asymmetry in the velocity
distribution, if the electron spot was asymmetric. For an ellipsoid with homogeneous charge density,
the velocity in the direction of the shorter axis is higher~\cite{Fluegge:Elektrodynamik:1986}. In
our case, the velocity distribution along $y$ direction was larger, as the size of the cloud is
smaller in this dimension. Simulating similar electron densities in smaller, but asymmetric volumes
showed an asymmetric velocity distribution as well. The velocity was higher in the direction of the
smaller expansion, corresponding to the $y$ direction in the experiment. Both effects resulted in
the vertical broadening of the electron pulse in velocity-mapping mode.

Simulations in cylindrical symmetry (ASTRA) for one million electrons per pulse provided an
approximate value for the pulse duration at the sample position of 60~ps. Albeit this was much
longer than in the case of a few electrons/pulse, it is sufficiently short for the diffractive
imaging of aligned and oriented molecules, which we can routinely create and control for hundreds of
picoseconds~\cite{Trippel:MP111:1738, Trippel:PRA89:051401R}.

\begin{figure}
   \centering
   \includegraphics[width=\linewidth]{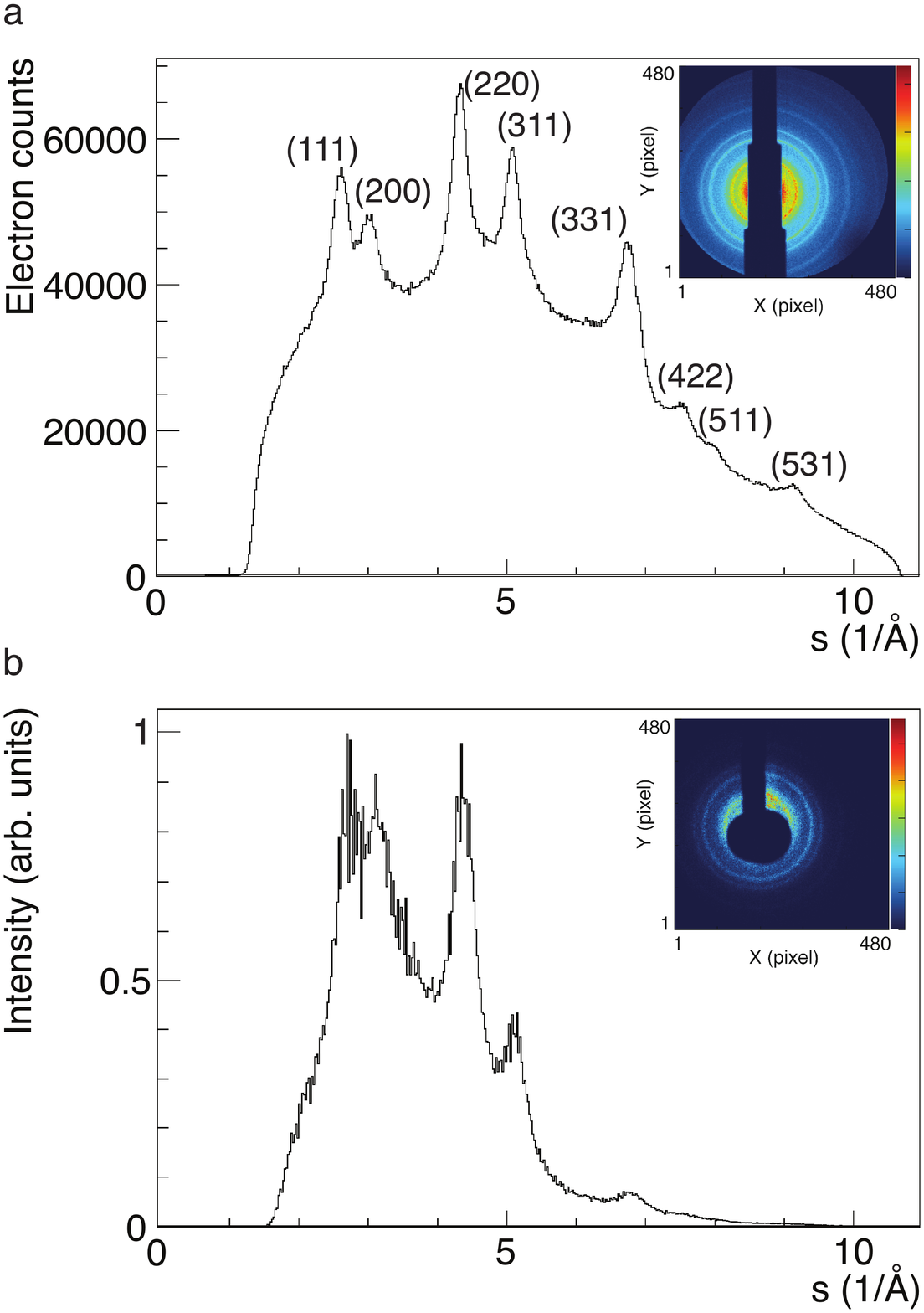}
   \caption{Radial scattering intensity for aluminum for (a) $10^{3}$ and (b) $10^{6}$ electrons per
      pulse. Peaks are labeled with Miller indices ($hkl$). The insets show the corresponding
      diffraction patterns.}
   \label{fig:diffraction}
\end{figure}
A thin polycrystalline aluminum sample was used to test the electron-optical properties of the
generated electron pulses, for instance, its coherence length and spatial resolution. The inset in
\autoref{fig:diffraction}\,a shows a diffraction pattern for $10^{3}$ electrons per pulse averaged
over $10^{6}$ pulses, \ie, about $15$~min at 1~kHz. The electron beam was focused on the detector,
which resulted in a nearly collimated beam at the position of the sample. The typical diffraction
rings of a polycrystalline sample were observed~\cite{Germer:PR56:58}. The corresponding radial
distribution as a function of momentum transfer $s$ is plotted in \autoref{fig:diffraction}\,a. The
peaks can be assigned to the allowed face-centered cubic crystal structure reflections for aluminum
and the corresponding Miller indices ($hkl$) are used to label the peaks~\cite{Germer:PR56:58}.

The inset in \autoref{fig:diffraction}\,b shows a diffraction pattern averaged over $10^3$ pulses
($\sim\!1$~s) with $10^{6}$ electrons per pulse. A 6-mm-wide beam stop was used, but the MCP voltage
still had to be reduced to avoid damaging the detector due to many electrons scattered to small $s$.
Reducing the MCP voltage reduces the gain of the detector system, \ie, the signal on the camera per
impinging electron. With this lower gain single, or a few, electrons could not be detected anymore.
Therefore, the peak intensities in \autoref{fig:diffraction}\,a and b cannot be compared
quantitatively. Diffraction peaks are still visible in \autoref{fig:diffraction}\,b, except for the
largest $s$ where the electron number and the gain were too small. This implies that the transverse
coherence of the electron pulses were larger than 234~pm, while approximately 1~nm was expected from
simulations. The spatial resolution of the imaging experiment was better than 234~pm, the
interatomic distance corresponding to the (111) reflection in the diffraction pattern of aluminum.
The restriction in resolution due to the lower detector gain will not occur in the envisioned
gas-phase experiments, as the sample density will be much smaller. Thus, single-electron detection
will also be possible for large electron numbers per pulse.

Using the implemented spectrometer it was possible to experimentally obtain the emittance, \ie, the
initial transverse spatial and velocity distributions of the electrons. The combination with
simulations allowed for the deduction of further values, such as the coherence length and pulse
duration of the propagated electron pulses. Compared to other sources with time resolutions on the
order of 1~ps or below~\cite{Ihee:Science291:458, Siwick:Science302:1382}, our setup did produce
electron pulses with 1.5~ps duration for few electrons/pulse. More importantly, our table-top setup
allows for a stable production of $>10^{6}$ electrons/pulse at a repetition rate of 1~kHz with an
estimated pulse durations of 60~ps. Nevertheless, due to the negligible cross-sections radiation
damage can be neglected even on these long timescales. For the prototypical 2,5-diiodobenzonitrile
molecule the effective electron-impact-ionization cross section is about $10^{-16}~\text{cm}^2$,
whereas the effective elastic-scattering cross section is $10^{-15}~\text{cm}^2$. For $10^{6}$
electrons per pulse, a molecular density of $10^{8}~\text{cm}^{-3}$, and an interaction volume of
$5\times10^{-6}~\text{cm}^3$ the signal on the detector of electrons elastically scattered off an
already destroyed molecule is
$S^D\approx0.5{\cdot}S^D_\text{elastic}{\cdot}S^D_\text{ionized}=0.5\cdot5\times10^{-3}\cdot5\times10^{-4}=1.25\times10^{-6}$
per shot~\cite{Kuepper:PRL112:083002}. This corresponds to a fraction of
$P^D=S^D\!/S^D_\text{elastic}=2.5\times10^{-4}$ electrons scattered off damaged molecules per
elastically-scattered signal electron. Thus, radiation damage is not relevant in these experiments.
Importantly, it is much smaller than in similar x-ray-diffraction experiments, where radiation
damage was not necessarily negligible, but could be reduced by increasing the x-ray beam
diameter~\cite{Kuepper:PRL112:083002}.

The current setup could be improved by increasing the acceleration fields: This would simultaneously
increase the electron number for the same laser power and lead to a smaller emittance and, thus, an
increased coherence length of the electron beam. The time resolution could either be improved
through a more compact design and stronger electric fields~\cite{Robinson:RSI86:013109} or by
combining the setup with an RF cavity for appropriate phase-space rotation for temporal
focusing~\cite{vanOudheusden:PRL105:264801, Fu:RIS85:083701, Sciaini:RPP74:096101}.

In conclusion, a new source for picosecond time-resolved electron diffraction experiments with the
need for large numbers of electrons was described. It will allow, for instance, the investigation of
dilute samples of controlled gas-phase molecules~\cite{Chang:IRPC:inprep}. Moreover, enabled by its
velocity map imaging spectrometer geometry, the setup allowed to characterize the electron beam
properties, \eg, the spatial and velocity distributions of the electrons. The focusing and coherence
properties of the electron pulses were determined through both, simulations and diffraction
experiments of aluminum-foil samples, to be sufficient for the envisioned atomically resolved
controlled molecule diffractive imaging.

We thank Sven Lederer, Ingo Hansen, and Hans-Hinrich Sahling for helpful advise regarding the
photocathode design and Klaus Floettmann for help with ASTRA simulations. We gratefully acknowledge
discussions with Martin Centurion, Jie Yang, Stephan Stern, and Henry N. Chapman on gas-phase
aligned-molecule diffraction as well as helpful discussions regarding electron diffraction with
Stuart Hayes, German Sciaini, Kostyantyn Pichugin, Julian Hirscht, and Dwayne Miller. We thank
Masaki Hada for the preparation of the aluminum sample.

Besides DESY, this work has been supported by the excellence cluster ``The Hamburg Center for
Ultrafast Imaging -- Structure, Dynamics and Control of Matter at the Atomic Scale'' of the Deutsche
Forschungsgemeinschaft (CUI, DFG-EXC1074) and the European Research Council through the Consolidator
Grant 614507-COMOTION. N.L.M.M. gratefully acknowledges a fellowship of the Joachim Herz Stiftung.

\bibliography{string,cmi}

\begin{thebibliography}{50}%
\makeatletter
\providecommand \@ifxundefined [1]{%
 \@ifx{#1\undefined}
}%
\providecommand \@ifnum [1]{%
 \ifnum #1\expandafter \@firstoftwo
 \else \expandafter \@secondoftwo
 \fi
}%
\providecommand \@ifx [1]{%
 \ifx #1\expandafter \@firstoftwo
 \else \expandafter \@secondoftwo
 \fi
}%
\providecommand \natexlab [1]{#1}%
\providecommand \enquote  [1]{``#1''}%
\providecommand \bibnamefont  [1]{#1}%
\providecommand \bibfnamefont [1]{#1}%
\providecommand \citenamefont [1]{#1}%
\providecommand \href@noop [0]{\@secondoftwo}%
\providecommand \href [0]{\begingroup \@sanitize@url \@href}%
\providecommand \@href[1]{\@@startlink{#1}\@@href}%
\providecommand \@@href[1]{\endgroup#1\@@endlink}%
\providecommand \@sanitize@url [0]{\catcode `\\12\catcode `\$12\catcode
  `\&12\catcode `\#12\catcode `\^12\catcode `\_12\catcode `\%12\relax}%
\providecommand \@@startlink[1]{}%
\providecommand \@@endlink[0]{}%
\providecommand \url  [0]{\begingroup\@sanitize@url \@url }%
\providecommand \@url [1]{\endgroup\@href {#1}{\urlprefix }}%
\providecommand \urlprefix  [0]{URL }%
\providecommand \Eprint [0]{\href }%
\providecommand \doibase [0]{http://dx.doi.org/}%
\providecommand \selectlanguage [0]{\@gobble}%
\providecommand \bibinfo  [0]{\@secondoftwo}%
\providecommand \bibfield  [0]{\@secondoftwo}%
\providecommand \translation [1]{[#1]}%
\providecommand \BibitemOpen [0]{}%
\providecommand \bibitemStop [0]{}%
\providecommand \bibitemNoStop [0]{.\EOS\space}%
\providecommand \EOS [0]{\spacefactor3000\relax}%
\providecommand \BibitemShut  [1]{\csname bibitem#1\endcsname}%
\let\auto@bib@innerbib\@empty
\bibitem [{\citenamefont {Germer}(1939)}]{Germer:PR56:58}%
  \BibitemOpen
  \bibfield  {author} {\bibinfo {author} {\bibfnamefont {L.}~\bibnamefont
  {Germer}},\ }\href {\doibase 10.1103/PhysRev.56.58} {\bibfield  {journal}
  {\bibinfo  {journal} {Phys.\ Rev.}\ }\textbf {\bibinfo {volume} {56}},\
  \bibinfo {pages} {58} (\bibinfo {year} {1939})}\BibitemShut {NoStop}%
\bibitem [{\citenamefont {Takayanagi}(1985)}]{Takayanagi:JVS3:1502}%
  \BibitemOpen
  \bibfield  {author} {\bibinfo {author} {\bibfnamefont {K.}~\bibnamefont
  {Takayanagi}},\ }\href {\doibase 10.1116/1.573160} {\bibfield  {journal}
  {\bibinfo  {journal} {J. Vac. Sci. Technol. A}\ }\textbf {\bibinfo {volume}
  {3}},\ \bibinfo {pages} {1502} (\bibinfo {year} {1985})}\BibitemShut
  {NoStop}%
\bibitem [{\citenamefont {Lonsdale}(1929)}]{Lonsdale:PRSA123:494}%
  \BibitemOpen
  \bibfield  {author} {\bibinfo {author} {\bibfnamefont {K.}~\bibnamefont
  {Lonsdale}},\ }\href {\doibase 10.1098/rspa.1929.0081} {\bibfield  {journal}
  {\bibinfo  {journal} {Proc. Royal Soc. London A}\ }\textbf {\bibinfo {volume}
  {123}},\ \bibinfo {pages} {494} (\bibinfo {year} {1929})}\BibitemShut
  {NoStop}%
\bibitem [{\citenamefont {Yonath}(2002)}]{Yonath:ARB1:257}%
  \BibitemOpen
  \bibfield  {author} {\bibinfo {author} {\bibfnamefont {A.}~\bibnamefont
  {Yonath}},\ }\href {\doibase 10.1146/annurev.biophys.31.082901.134439}
  {\bibfield  {journal} {\bibinfo  {journal} {Annu. Rev. Biophys. Biomol.
  Struct.}\ }\textbf {\bibinfo {volume} {31}},\ \bibinfo {pages} {257}
  (\bibinfo {year} {2002})}\BibitemShut {NoStop}%
\bibitem [{\citenamefont {Hargittai}\ and\ \citenamefont
  {Hargittai}(1988)}]{Hargittai:GED:1988}%
  \BibitemOpen
  \bibfield  {author} {\bibinfo {author} {\bibfnamefont {I.}~\bibnamefont
  {Hargittai}}\ and\ \bibinfo {author} {\bibfnamefont {M.}~\bibnamefont
  {Hargittai}},\ }\href
  {http://www.wiley-vch.de/publish/en/books/ISBN978-0-471-18689-2} {\emph
  {\bibinfo {title} {Stereochemical Applications of Gas-Phase Electron
  Diffraction}}}\ (\bibinfo  {publisher} {VCH Verlagsgesellschaft},\ \bibinfo
  {year} {1988})\BibitemShut {NoStop}%
\bibitem [{\citenamefont {Williamson}\ \emph {et~al.}(1997)\citenamefont
  {Williamson}, \citenamefont {Cao}, \citenamefont {Ihee}, \citenamefont
  {Frey},\ and\ \citenamefont {Zewail}}]{Williamson:Nature386:159}%
  \BibitemOpen
  \bibfield  {author} {\bibinfo {author} {\bibfnamefont {J.~C.}\ \bibnamefont
  {Williamson}}, \bibinfo {author} {\bibfnamefont {J.~M.}\ \bibnamefont {Cao}},
  \bibinfo {author} {\bibfnamefont {H.}~\bibnamefont {Ihee}}, \bibinfo {author}
  {\bibfnamefont {H.}~\bibnamefont {Frey}}, \ and\ \bibinfo {author}
  {\bibfnamefont {A.~H.}\ \bibnamefont {Zewail}},\ }\href {\doibase
  10.1038/386159a0} {\bibfield  {journal} {\bibinfo  {journal} {Nature}\
  }\textbf {\bibinfo {volume} {386}},\ \bibinfo {pages} {159} (\bibinfo {year}
  {1997})}\BibitemShut {NoStop}%
\bibitem [{\citenamefont {Ihee}\ \emph {et~al.}(2001)\citenamefont {Ihee},
  \citenamefont {Lobastov}, \citenamefont {Gomez}, \citenamefont {Goodson},
  \citenamefont {Srinivasan}, \citenamefont {Ruan},\ and\ \citenamefont
  {Zewail}}]{Ihee:Science291:458}%
  \BibitemOpen
  \bibfield  {author} {\bibinfo {author} {\bibfnamefont {H.}~\bibnamefont
  {Ihee}}, \bibinfo {author} {\bibfnamefont {V.}~\bibnamefont {Lobastov}},
  \bibinfo {author} {\bibfnamefont {U.}~\bibnamefont {Gomez}}, \bibinfo
  {author} {\bibfnamefont {B.}~\bibnamefont {Goodson}}, \bibinfo {author}
  {\bibfnamefont {R.}~\bibnamefont {Srinivasan}}, \bibinfo {author}
  {\bibfnamefont {C.}~\bibnamefont {Ruan}}, \ and\ \bibinfo {author}
  {\bibfnamefont {A.~H.}\ \bibnamefont {Zewail}},\ }\href {\doibase
  10.1126/science.291.5503.458} {\bibfield  {journal} {\bibinfo  {journal}
  {Science}\ }\textbf {\bibinfo {volume} {291}},\ \bibinfo {pages} {458}
  (\bibinfo {year} {2001})}\BibitemShut {NoStop}%
\bibitem [{\citenamefont {Siwick}\ \emph {et~al.}(2003)\citenamefont {Siwick},
  \citenamefont {Dwyer}, \citenamefont {Jordan},\ and\ \citenamefont
  {Miller}}]{Siwick:Science302:1382}%
  \BibitemOpen
  \bibfield  {author} {\bibinfo {author} {\bibfnamefont {B.~J.}\ \bibnamefont
  {Siwick}}, \bibinfo {author} {\bibfnamefont {J.~R.}\ \bibnamefont {Dwyer}},
  \bibinfo {author} {\bibfnamefont {R.~E.}\ \bibnamefont {Jordan}}, \ and\
  \bibinfo {author} {\bibfnamefont {R.~J.~D.}\ \bibnamefont {Miller}},\ }\href
  {\doibase 10.1126/science.1090052} {\bibfield  {journal} {\bibinfo  {journal}
  {Science}\ }\textbf {\bibinfo {volume} {302}},\ \bibinfo {pages} {1382}
  (\bibinfo {year} {2003})}\BibitemShut {NoStop}%
\bibitem [{\citenamefont {Chapman}\ \emph {et~al.}(2006)\citenamefont
  {Chapman}, \citenamefont {Barty}, \citenamefont {Bogan}, \citenamefont
  {Boutet}, \citenamefont {Frank}, \citenamefont {Hau-Riege}, \citenamefont
  {Marchesini}, \citenamefont {Woods}, \citenamefont {Bajt}, \citenamefont
  {Benner}, \citenamefont {A.}, \citenamefont {Pl{\"o}njes}, \citenamefont
  {Kuhlmann}, \citenamefont {Treusch}, \citenamefont {D{\"u}sterer},
  \citenamefont {Tschentscher}, \citenamefont {Schneider}, \citenamefont
  {Spiller}, \citenamefont {M{\"o}ller}, \citenamefont {Bostedt}, \citenamefont
  {Hoener}, \citenamefont {Shapiro}, \citenamefont {Hodgson}, \citenamefont
  {van~der Spoel}, \citenamefont {Burmeister}, \citenamefont {Bergh},
  \citenamefont {Caleman}, \citenamefont {Huldt}, \citenamefont {Seibert},
  \citenamefont {Maia}, \citenamefont {Lee}, \citenamefont {Sz{\"o}ke},
  \citenamefont {Timneanu},\ and\ \citenamefont
  {Hajdu}}]{Chapman:NatPhys2:839}%
  \BibitemOpen
  \bibfield  {author} {\bibinfo {author} {\bibfnamefont {H.~N.}\ \bibnamefont
  {Chapman}}, \bibinfo {author} {\bibfnamefont {A.}~\bibnamefont {Barty}},
  \bibinfo {author} {\bibfnamefont {M.~J.}\ \bibnamefont {Bogan}}, \bibinfo
  {author} {\bibfnamefont {S.}~\bibnamefont {Boutet}}, \bibinfo {author}
  {\bibfnamefont {S.}~\bibnamefont {Frank}}, \bibinfo {author} {\bibfnamefont
  {S.~P.}\ \bibnamefont {Hau-Riege}}, \bibinfo {author} {\bibfnamefont
  {S.}~\bibnamefont {Marchesini}}, \bibinfo {author} {\bibfnamefont {B.~W.}\
  \bibnamefont {Woods}}, \bibinfo {author} {\bibfnamefont {S.}~\bibnamefont
  {Bajt}}, \bibinfo {author} {\bibfnamefont {W.~H.}\ \bibnamefont {Benner}},
  \bibinfo {author} {\bibfnamefont {L.~W.}\ \bibnamefont {A.}}, \bibinfo
  {author} {\bibfnamefont {E.}~\bibnamefont {Pl{\"o}njes}}, \bibinfo {author}
  {\bibfnamefont {M.}~\bibnamefont {Kuhlmann}}, \bibinfo {author}
  {\bibfnamefont {R.}~\bibnamefont {Treusch}}, \bibinfo {author} {\bibfnamefont
  {S.}~\bibnamefont {D{\"u}sterer}}, \bibinfo {author} {\bibfnamefont
  {T.}~\bibnamefont {Tschentscher}}, \bibinfo {author} {\bibfnamefont {J.~R.}\
  \bibnamefont {Schneider}}, \bibinfo {author} {\bibfnamefont {E.}~\bibnamefont
  {Spiller}}, \bibinfo {author} {\bibfnamefont {T.}~\bibnamefont {M{\"o}ller}},
  \bibinfo {author} {\bibfnamefont {C.}~\bibnamefont {Bostedt}}, \bibinfo
  {author} {\bibfnamefont {M.}~\bibnamefont {Hoener}}, \bibinfo {author}
  {\bibfnamefont {D.~A.}\ \bibnamefont {Shapiro}}, \bibinfo {author}
  {\bibfnamefont {K.~O.}\ \bibnamefont {Hodgson}}, \bibinfo {author}
  {\bibfnamefont {D.}~\bibnamefont {van~der Spoel}}, \bibinfo {author}
  {\bibfnamefont {F.}~\bibnamefont {Burmeister}}, \bibinfo {author}
  {\bibfnamefont {M.}~\bibnamefont {Bergh}}, \bibinfo {author} {\bibfnamefont
  {C.}~\bibnamefont {Caleman}}, \bibinfo {author} {\bibfnamefont
  {G.}~\bibnamefont {Huldt}}, \bibinfo {author} {\bibfnamefont {M.~M.}\
  \bibnamefont {Seibert}}, \bibinfo {author} {\bibfnamefont {F.~R. N.~C.}\
  \bibnamefont {Maia}}, \bibinfo {author} {\bibfnamefont {R.~W.}\ \bibnamefont
  {Lee}}, \bibinfo {author} {\bibfnamefont {A.}~\bibnamefont {Sz{\"o}ke}},
  \bibinfo {author} {\bibfnamefont {N.}~\bibnamefont {Timneanu}}, \ and\
  \bibinfo {author} {\bibfnamefont {J.}~\bibnamefont {Hajdu}},\ }\href
  {\doibase 10.1038/nphys461} {\bibfield  {journal} {\bibinfo  {journal} {Nat.
  Phys.}\ }\textbf {\bibinfo {volume} {2}},\ \bibinfo {pages} {839} (\bibinfo
  {year} {2006})}\BibitemShut {NoStop}%
\bibitem [{\citenamefont {Spence}\ \emph {et~al.}(2012)\citenamefont {Spence},
  \citenamefont {Weierstall},\ and\ \citenamefont
  {Chapman}}]{Spence:RPP75:102601}%
  \BibitemOpen
  \bibfield  {author} {\bibinfo {author} {\bibfnamefont {J.~C.~H.}\
  \bibnamefont {Spence}}, \bibinfo {author} {\bibfnamefont {U.}~\bibnamefont
  {Weierstall}}, \ and\ \bibinfo {author} {\bibfnamefont {H.~N.}\ \bibnamefont
  {Chapman}},\ }\href {\doibase 10.1088/0034-4885/75/10/102601} {\bibfield
  {journal} {\bibinfo  {journal} {Rep.\ Prog.\ Phys.}\ }\textbf {\bibinfo
  {volume} {75}},\ \bibinfo {pages} {102601} (\bibinfo {year}
  {2012})}\BibitemShut {NoStop}%
\bibitem [{\citenamefont {Barty}\ \emph {et~al.}(2013)\citenamefont {Barty},
  \citenamefont {K{\"u}pper},\ and\ \citenamefont
  {Chapman}}]{Barty:ARPC64:415}%
  \BibitemOpen
  \bibfield  {author} {\bibinfo {author} {\bibfnamefont {A.}~\bibnamefont
  {Barty}}, \bibinfo {author} {\bibfnamefont {J.}~\bibnamefont {K{\"u}pper}}, \
  and\ \bibinfo {author} {\bibfnamefont {H.~N.}\ \bibnamefont {Chapman}},\
  }\href {\doibase 10.1146/annurev-physchem-032511-143708} {\bibfield
  {journal} {\bibinfo  {journal} {Annu.\ Rev.\ Phys.\ Chem.}\ }\textbf
  {\bibinfo {volume} {64}},\ \bibinfo {pages} {415} (\bibinfo {year}
  {2013})}\BibitemShut {NoStop}%
\bibitem [{\citenamefont {Chang}\ \emph {et~al.}(2015)\citenamefont {Chang},
  \citenamefont {Horke}, \citenamefont {Trippel},\ and\ \citenamefont
  {Küpper}}]{Chang:IRPC:inprep}%
  \BibitemOpen
  \bibfield  {author} {\bibinfo {author} {\bibfnamefont {Y.-P.}\ \bibnamefont
  {Chang}}, \bibinfo {author} {\bibfnamefont {D.}~\bibnamefont {Horke}},
  \bibinfo {author} {\bibfnamefont {S.}~\bibnamefont {Trippel}}, \ and\
  \bibinfo {author} {\bibfnamefont {J.}~\bibnamefont {Küpper}},\ }\href
  {\doibase 10.1080/0144235X.2015.1077838} {\bibfield  {journal} {\bibinfo
  {journal} {Int.\ Rev.\ Phys.\ Chem.}\ } (\bibinfo {year} {2015}),\
  10.1080/0144235X.2015.1077838},\ \bibinfo {note} {in press},\ \Eprint
  {http://arxiv.org/abs/1505.05632} {arXiv:1505.05632 [physics]} \BibitemShut
  {NoStop}%
\bibitem [{\citenamefont {Filsinger}\ \emph {et~al.}(2009)\citenamefont
  {Filsinger}, \citenamefont {K{\"u}pper}, \citenamefont {Meijer},
  \citenamefont {Hansen}, \citenamefont {Maurer}, \citenamefont {Nielsen},
  \citenamefont {Holmegaard},\ and\ \citenamefont
  {Stapelfeldt}}]{Filsinger:ACIE48:6900}%
  \BibitemOpen
  \bibfield  {author} {\bibinfo {author} {\bibfnamefont {F.}~\bibnamefont
  {Filsinger}}, \bibinfo {author} {\bibfnamefont {J.}~\bibnamefont
  {K{\"u}pper}}, \bibinfo {author} {\bibfnamefont {G.}~\bibnamefont {Meijer}},
  \bibinfo {author} {\bibfnamefont {J.~L.}\ \bibnamefont {Hansen}}, \bibinfo
  {author} {\bibfnamefont {J.}~\bibnamefont {Maurer}}, \bibinfo {author}
  {\bibfnamefont {J.~H.}\ \bibnamefont {Nielsen}}, \bibinfo {author}
  {\bibfnamefont {L.}~\bibnamefont {Holmegaard}}, \ and\ \bibinfo {author}
  {\bibfnamefont {H.}~\bibnamefont {Stapelfeldt}},\ }\href {\doibase
  10.1002/anie.200902650} {\bibfield  {journal} {\bibinfo  {journal} {Angew.\
  Chem.\ Int.\ Ed.}\ }\textbf {\bibinfo {volume} {48}},\ \bibinfo {pages}
  {6900} (\bibinfo {year} {2009})}\BibitemShut {NoStop}%
\bibitem [{\citenamefont {Kierspel}\ \emph {et~al.}(2014)\citenamefont
  {Kierspel}, \citenamefont {Horke}, \citenamefont {Chang},\ and\ \citenamefont
  {K{\"u}pper}}]{Kierspel:CPL591:130}%
  \BibitemOpen
  \bibfield  {author} {\bibinfo {author} {\bibfnamefont {T.}~\bibnamefont
  {Kierspel}}, \bibinfo {author} {\bibfnamefont {D.~A.}\ \bibnamefont {Horke}},
  \bibinfo {author} {\bibfnamefont {Y.-P.}\ \bibnamefont {Chang}}, \ and\
  \bibinfo {author} {\bibfnamefont {J.}~\bibnamefont {K{\"u}pper}},\ }\href
  {\doibase 10.1016/j.cplett.2013.11.010} {\bibfield  {journal} {\bibinfo
  {journal} {Chem.\ Phys.\ Lett.}\ }\textbf {\bibinfo {volume} {591}},\
  \bibinfo {pages} {130} (\bibinfo {year} {2014})},\ \Eprint
  {http://arxiv.org/abs/1312.4417} {arXiv:1312.4417 [physics]} \BibitemShut
  {NoStop}%
\bibitem [{\citenamefont {Filsinger}\ \emph {et~al.}(2008)\citenamefont
  {Filsinger}, \citenamefont {Erlekam}, \citenamefont {von Helden},
  \citenamefont {K{\"u}pper},\ and\ \citenamefont
  {Meijer}}]{Filsinger:PRL100:133003}%
  \BibitemOpen
  \bibfield  {author} {\bibinfo {author} {\bibfnamefont {F.}~\bibnamefont
  {Filsinger}}, \bibinfo {author} {\bibfnamefont {U.}~\bibnamefont {Erlekam}},
  \bibinfo {author} {\bibfnamefont {G.}~\bibnamefont {von Helden}}, \bibinfo
  {author} {\bibfnamefont {J.}~\bibnamefont {K{\"u}pper}}, \ and\ \bibinfo
  {author} {\bibfnamefont {G.}~\bibnamefont {Meijer}},\ }\href {\doibase
  10.1103/PhysRevLett.100.133003} {\bibfield  {journal} {\bibinfo  {journal}
  {Phys.\ Rev.\ Lett.}\ }\textbf {\bibinfo {volume} {100}},\ \bibinfo {pages}
  {133003} (\bibinfo {year} {2008})},\ \Eprint {http://arxiv.org/abs/0802.2795}
  {arXiv:0802.2795 [physics]} \BibitemShut {NoStop}%
\bibitem [{\citenamefont {Trippel}\ \emph {et~al.}(2012)\citenamefont
  {Trippel}, \citenamefont {Chang}, \citenamefont {Stern}, \citenamefont
  {Mullins}, \citenamefont {Holmegaard},\ and\ \citenamefont
  {K{\"u}pper}}]{Trippel:PRA86:033202}%
  \BibitemOpen
  \bibfield  {author} {\bibinfo {author} {\bibfnamefont {S.}~\bibnamefont
  {Trippel}}, \bibinfo {author} {\bibfnamefont {Y.-P.}\ \bibnamefont {Chang}},
  \bibinfo {author} {\bibfnamefont {S.}~\bibnamefont {Stern}}, \bibinfo
  {author} {\bibfnamefont {T.}~\bibnamefont {Mullins}}, \bibinfo {author}
  {\bibfnamefont {L.}~\bibnamefont {Holmegaard}}, \ and\ \bibinfo {author}
  {\bibfnamefont {J.}~\bibnamefont {K{\"u}pper}},\ }\href {\doibase
  10.1103/PhysRevA.86.033202} {\bibfield  {journal} {\bibinfo  {journal}
  {Phys.\ Rev.\ A}\ }\textbf {\bibinfo {volume} {86}},\ \bibinfo {pages}
  {033202} (\bibinfo {year} {2012})},\ \Eprint {http://arxiv.org/abs/1208.4935}
  {arXiv:1208.4935 [physics]} \BibitemShut {NoStop}%
\bibitem [{\citenamefont {Ryu}\ \emph {et~al.}(2003)\citenamefont {Ryu},
  \citenamefont {Stratt},\ and\ \citenamefont {Weber}}]{Ryu:JPCA107:6622}%
  \BibitemOpen
  \bibfield  {author} {\bibinfo {author} {\bibfnamefont {S.}~\bibnamefont
  {Ryu}}, \bibinfo {author} {\bibfnamefont {R.}~\bibnamefont {Stratt}}, \ and\
  \bibinfo {author} {\bibfnamefont {P.}~\bibnamefont {Weber}},\ }\href
  {\doibase 10.1021/jp0304632} {\bibfield  {journal} {\bibinfo  {journal} {J.\
  Phys.\ Chem.\ A}\ }\textbf {\bibinfo {volume} {107}},\ \bibinfo {pages}
  {6622} (\bibinfo {year} {2003})}\BibitemShut {NoStop}%
\bibitem [{\citenamefont {Spence}\ and\ \citenamefont
  {Doak}(2004)}]{Spence:PRL92:198102}%
  \BibitemOpen
  \bibfield  {author} {\bibinfo {author} {\bibfnamefont {J.~C.~H.}\
  \bibnamefont {Spence}}\ and\ \bibinfo {author} {\bibfnamefont {R.~B.}\
  \bibnamefont {Doak}},\ }\href {\doibase 10.1103/PhysRevLett.92.198102}
  {\bibfield  {journal} {\bibinfo  {journal} {Phys.\ Rev.\ Lett.}\ }\textbf
  {\bibinfo {volume} {92}},\ \bibinfo {pages} {198102} (\bibinfo {year}
  {2004})}\BibitemShut {NoStop}%
\bibitem [{\citenamefont {Pabst}\ \emph {et~al.}(2010)\citenamefont {Pabst},
  \citenamefont {Ho},\ and\ \citenamefont {Santra}}]{Pabst:PRA81:043425}%
  \BibitemOpen
  \bibfield  {author} {\bibinfo {author} {\bibfnamefont {S.}~\bibnamefont
  {Pabst}}, \bibinfo {author} {\bibfnamefont {P.~J.}\ \bibnamefont {Ho}}, \
  and\ \bibinfo {author} {\bibfnamefont {R.}~\bibnamefont {Santra}},\ }\href
  {\doibase 10.1103/PhysRevA.81.043425} {\bibfield  {journal} {\bibinfo
  {journal} {Phys.\ Rev.\ A}\ }\textbf {\bibinfo {volume} {81}},\ \bibinfo
  {pages} {043425} (\bibinfo {year} {2010})}\BibitemShut {NoStop}%
\bibitem [{\citenamefont {Filsinger}\ \emph {et~al.}(2011)\citenamefont
  {Filsinger}, \citenamefont {Meijer}, \citenamefont {Stapelfeldt},
  \citenamefont {Chapman},\ and\ \citenamefont
  {K{\"u}pper}}]{Filsinger:PCCP13:2076}%
  \BibitemOpen
  \bibfield  {author} {\bibinfo {author} {\bibfnamefont {F.}~\bibnamefont
  {Filsinger}}, \bibinfo {author} {\bibfnamefont {G.}~\bibnamefont {Meijer}},
  \bibinfo {author} {\bibfnamefont {H.}~\bibnamefont {Stapelfeldt}}, \bibinfo
  {author} {\bibfnamefont {H.}~\bibnamefont {Chapman}}, \ and\ \bibinfo
  {author} {\bibfnamefont {J.}~\bibnamefont {K{\"u}pper}},\ }\href {\doibase
  10.1039/C0CP01585G} {\bibfield  {journal} {\bibinfo  {journal} {Phys.\ Chem.\
  Chem.\ Phys.}\ }\textbf {\bibinfo {volume} {13}},\ \bibinfo {pages} {2076}
  (\bibinfo {year} {2011})}\BibitemShut {NoStop}%
\bibitem [{\citenamefont {K{\"u}pper}\ \emph {et~al.}(2014)\citenamefont
  {K{\"u}pper}, \citenamefont {Stern}, \citenamefont {Holmegaard},
  \citenamefont {Filsinger}, \citenamefont {Rouz\'{e}e}, \citenamefont
  {Rudenko}, \citenamefont {Johnsson}, \citenamefont {Martin}, \citenamefont
  {Adolph}, \citenamefont {Aquila}, \citenamefont {Bajt}, \citenamefont
  {Barty}, \citenamefont {Bostedt}, \citenamefont {Bozek}, \citenamefont
  {Caleman}, \citenamefont {Coffee}, \citenamefont {Coppola}, \citenamefont
  {Delmas}, \citenamefont {Epp}, \citenamefont {Erk}, \citenamefont {Foucar},
  \citenamefont {Gorkhover}, \citenamefont {Gumprecht}, \citenamefont
  {Hartmann}, \citenamefont {Hartmann}, \citenamefont {Hauser}, \citenamefont
  {Holl}, \citenamefont {H{\"o}mke}, \citenamefont {Kimmel}, \citenamefont
  {Krasniqi}, \citenamefont {K{\"u}hnel}, \citenamefont {Maurer}, \citenamefont
  {Messerschmidt}, \citenamefont {Moshammer}, \citenamefont {Reich},
  \citenamefont {Rudek}, \citenamefont {Santra}, \citenamefont {Schlichting},
  \citenamefont {Schmidt}, \citenamefont {Schorb}, \citenamefont {Schulz},
  \citenamefont {Soltau}, \citenamefont {Spence}, \citenamefont {Starodub},
  \citenamefont {Str{\"u}der}, \citenamefont {Th{\o}gersen}, \citenamefont
  {Vrakking}, \citenamefont {Weidenspointner}, \citenamefont {White},
  \citenamefont {Wunderer}, \citenamefont {Meijer}, \citenamefont {Ullrich},
  \citenamefont {Stapelfeldt}, \citenamefont {Rolles},\ and\ \citenamefont
  {Chapman}}]{Kuepper:PRL112:083002}%
  \BibitemOpen
  \bibfield  {author} {\bibinfo {author} {\bibfnamefont {J.}~\bibnamefont
  {K{\"u}pper}}, \bibinfo {author} {\bibfnamefont {S.}~\bibnamefont {Stern}},
  \bibinfo {author} {\bibfnamefont {L.}~\bibnamefont {Holmegaard}}, \bibinfo
  {author} {\bibfnamefont {F.}~\bibnamefont {Filsinger}}, \bibinfo {author}
  {\bibfnamefont {A.}~\bibnamefont {Rouz\'{e}e}}, \bibinfo {author}
  {\bibfnamefont {A.}~\bibnamefont {Rudenko}}, \bibinfo {author} {\bibfnamefont
  {P.}~\bibnamefont {Johnsson}}, \bibinfo {author} {\bibfnamefont {A.~V.}\
  \bibnamefont {Martin}}, \bibinfo {author} {\bibfnamefont {M.}~\bibnamefont
  {Adolph}}, \bibinfo {author} {\bibfnamefont {A.}~\bibnamefont {Aquila}},
  \bibinfo {author} {\bibfnamefont {S.}~\bibnamefont {Bajt}}, \bibinfo {author}
  {\bibfnamefont {A.}~\bibnamefont {Barty}}, \bibinfo {author} {\bibfnamefont
  {C.}~\bibnamefont {Bostedt}}, \bibinfo {author} {\bibfnamefont
  {J.}~\bibnamefont {Bozek}}, \bibinfo {author} {\bibfnamefont
  {C.}~\bibnamefont {Caleman}}, \bibinfo {author} {\bibfnamefont
  {R.}~\bibnamefont {Coffee}}, \bibinfo {author} {\bibfnamefont
  {N.}~\bibnamefont {Coppola}}, \bibinfo {author} {\bibfnamefont
  {T.}~\bibnamefont {Delmas}}, \bibinfo {author} {\bibfnamefont
  {S.}~\bibnamefont {Epp}}, \bibinfo {author} {\bibfnamefont {B.}~\bibnamefont
  {Erk}}, \bibinfo {author} {\bibfnamefont {L.}~\bibnamefont {Foucar}},
  \bibinfo {author} {\bibfnamefont {T.}~\bibnamefont {Gorkhover}}, \bibinfo
  {author} {\bibfnamefont {L.}~\bibnamefont {Gumprecht}}, \bibinfo {author}
  {\bibfnamefont {A.}~\bibnamefont {Hartmann}}, \bibinfo {author}
  {\bibfnamefont {R.}~\bibnamefont {Hartmann}}, \bibinfo {author}
  {\bibfnamefont {G.}~\bibnamefont {Hauser}}, \bibinfo {author} {\bibfnamefont
  {P.}~\bibnamefont {Holl}}, \bibinfo {author} {\bibfnamefont {A.}~\bibnamefont
  {H{\"o}mke}}, \bibinfo {author} {\bibfnamefont {N.}~\bibnamefont {Kimmel}},
  \bibinfo {author} {\bibfnamefont {F.}~\bibnamefont {Krasniqi}}, \bibinfo
  {author} {\bibfnamefont {K.-U.}\ \bibnamefont {K{\"u}hnel}}, \bibinfo
  {author} {\bibfnamefont {J.}~\bibnamefont {Maurer}}, \bibinfo {author}
  {\bibfnamefont {M.}~\bibnamefont {Messerschmidt}}, \bibinfo {author}
  {\bibfnamefont {R.}~\bibnamefont {Moshammer}}, \bibinfo {author}
  {\bibfnamefont {C.}~\bibnamefont {Reich}}, \bibinfo {author} {\bibfnamefont
  {B.}~\bibnamefont {Rudek}}, \bibinfo {author} {\bibfnamefont
  {R.}~\bibnamefont {Santra}}, \bibinfo {author} {\bibfnamefont
  {I.}~\bibnamefont {Schlichting}}, \bibinfo {author} {\bibfnamefont
  {C.}~\bibnamefont {Schmidt}}, \bibinfo {author} {\bibfnamefont
  {S.}~\bibnamefont {Schorb}}, \bibinfo {author} {\bibfnamefont
  {J.}~\bibnamefont {Schulz}}, \bibinfo {author} {\bibfnamefont
  {H.}~\bibnamefont {Soltau}}, \bibinfo {author} {\bibfnamefont {J.~C.~H.}\
  \bibnamefont {Spence}}, \bibinfo {author} {\bibfnamefont {D.}~\bibnamefont
  {Starodub}}, \bibinfo {author} {\bibfnamefont {L.}~\bibnamefont
  {Str{\"u}der}}, \bibinfo {author} {\bibfnamefont {J.}~\bibnamefont
  {Th{\o}gersen}}, \bibinfo {author} {\bibfnamefont {M.~J.~J.}\ \bibnamefont
  {Vrakking}}, \bibinfo {author} {\bibfnamefont {G.}~\bibnamefont
  {Weidenspointner}}, \bibinfo {author} {\bibfnamefont {T.~A.}\ \bibnamefont
  {White}}, \bibinfo {author} {\bibfnamefont {C.}~\bibnamefont {Wunderer}},
  \bibinfo {author} {\bibfnamefont {G.}~\bibnamefont {Meijer}}, \bibinfo
  {author} {\bibfnamefont {J.}~\bibnamefont {Ullrich}}, \bibinfo {author}
  {\bibfnamefont {H.}~\bibnamefont {Stapelfeldt}}, \bibinfo {author}
  {\bibfnamefont {D.}~\bibnamefont {Rolles}}, \ and\ \bibinfo {author}
  {\bibfnamefont {H.~N.}\ \bibnamefont {Chapman}},\ }\href {\doibase
  10.1103/PhysRevLett.112.083002} {\bibfield  {journal} {\bibinfo  {journal}
  {Phys.\ Rev.\ Lett.}\ }\textbf {\bibinfo {volume} {112}},\ \bibinfo {pages}
  {083002} (\bibinfo {year} {2014})},\ \Eprint {http://arxiv.org/abs/1307.4577}
  {arXiv:1307.4577 [physics]} \BibitemShut {NoStop}%
\bibitem [{\citenamefont {Stern}\ \emph {et~al.}(2014)\citenamefont {Stern},
  \citenamefont {Holmegaard}, \citenamefont {Filsinger}, \citenamefont
  {Rouzee}, \citenamefont {Rudenko}, \citenamefont {Johnsson}, \citenamefont
  {Martin}, \citenamefont {Barty}, \citenamefont {Bostedt}, \citenamefont
  {Bozek}, \citenamefont {Coffee}, \citenamefont {Epp}, \citenamefont {Erk},
  \citenamefont {Foucar}, \citenamefont {Hartmann}, \citenamefont {Kimmel},
  \citenamefont {Kühnel}, \citenamefont {Maurer}, \citenamefont
  {Messerschmidt}, \citenamefont {Rudek}, \citenamefont {Starodub},
  \citenamefont {Thøgersen}, \citenamefont {Weidenspointner}, \citenamefont
  {White}, \citenamefont {Stapelfeldt}, \citenamefont {Rolles}, \citenamefont
  {Chapman},\ and\ \citenamefont {Küpper}}]{Stern:FD171:393}%
  \BibitemOpen
  \bibfield  {author} {\bibinfo {author} {\bibfnamefont {S.}~\bibnamefont
  {Stern}}, \bibinfo {author} {\bibfnamefont {L.}~\bibnamefont {Holmegaard}},
  \bibinfo {author} {\bibfnamefont {F.}~\bibnamefont {Filsinger}}, \bibinfo
  {author} {\bibfnamefont {A.}~\bibnamefont {Rouzee}}, \bibinfo {author}
  {\bibfnamefont {A.}~\bibnamefont {Rudenko}}, \bibinfo {author} {\bibfnamefont
  {P.}~\bibnamefont {Johnsson}}, \bibinfo {author} {\bibfnamefont {A.~V.}\
  \bibnamefont {Martin}}, \bibinfo {author} {\bibfnamefont {A.}~\bibnamefont
  {Barty}}, \bibinfo {author} {\bibfnamefont {C.}~\bibnamefont {Bostedt}},
  \bibinfo {author} {\bibfnamefont {J.}~\bibnamefont {Bozek}}, \bibinfo
  {author} {\bibfnamefont {R.}~\bibnamefont {Coffee}}, \bibinfo {author}
  {\bibfnamefont {S.}~\bibnamefont {Epp}}, \bibinfo {author} {\bibfnamefont
  {B.}~\bibnamefont {Erk}}, \bibinfo {author} {\bibfnamefont {L.}~\bibnamefont
  {Foucar}}, \bibinfo {author} {\bibfnamefont {R.}~\bibnamefont {Hartmann}},
  \bibinfo {author} {\bibfnamefont {N.}~\bibnamefont {Kimmel}}, \bibinfo
  {author} {\bibfnamefont {K.-U.}\ \bibnamefont {Kühnel}}, \bibinfo {author}
  {\bibfnamefont {J.}~\bibnamefont {Maurer}}, \bibinfo {author} {\bibfnamefont
  {M.}~\bibnamefont {Messerschmidt}}, \bibinfo {author} {\bibfnamefont
  {B.}~\bibnamefont {Rudek}}, \bibinfo {author} {\bibfnamefont
  {D.}~\bibnamefont {Starodub}}, \bibinfo {author} {\bibfnamefont
  {J.}~\bibnamefont {Thøgersen}}, \bibinfo {author} {\bibfnamefont
  {G.}~\bibnamefont {Weidenspointner}}, \bibinfo {author} {\bibfnamefont
  {T.~A.}\ \bibnamefont {White}}, \bibinfo {author} {\bibfnamefont
  {H.}~\bibnamefont {Stapelfeldt}}, \bibinfo {author} {\bibfnamefont
  {D.}~\bibnamefont {Rolles}}, \bibinfo {author} {\bibfnamefont {H.~N.}\
  \bibnamefont {Chapman}}, \ and\ \bibinfo {author} {\bibfnamefont
  {J.}~\bibnamefont {Küpper}},\ }\href {\doibase 10.1039/c4fd00028e}
  {\bibfield  {journal} {\bibinfo  {journal} {Faraday Disc.}\ }\textbf
  {\bibinfo {volume} {171}},\ \bibinfo {pages} {393} (\bibinfo {year}
  {2014})},\ \Eprint {http://arxiv.org/abs/1403.2553} {arXiv:1403.2553
  [physics]} \BibitemShut {NoStop}%
\bibitem [{\citenamefont {Hensley}\ \emph {et~al.}(2012)\citenamefont
  {Hensley}, \citenamefont {Yang},\ and\ \citenamefont
  {Centurion}}]{Hensley:PRL109:133202}%
  \BibitemOpen
  \bibfield  {author} {\bibinfo {author} {\bibfnamefont {C.~J.}\ \bibnamefont
  {Hensley}}, \bibinfo {author} {\bibfnamefont {J.}~\bibnamefont {Yang}}, \
  and\ \bibinfo {author} {\bibfnamefont {M.}~\bibnamefont {Centurion}},\ }\href
  {\doibase 10.1103/PhysRevLett.109.133202} {\bibfield  {journal} {\bibinfo
  {journal} {Phys.\ Rev.\ Lett.}\ }\textbf {\bibinfo {volume} {109}},\ \bibinfo
  {pages} {133202} (\bibinfo {year} {2012})}\BibitemShut {NoStop}%
\bibitem [{\citenamefont {Yang}\ \emph {et~al.}(2014)\citenamefont {Yang},
  \citenamefont {Makhija}, \citenamefont {Kumarappan},\ and\ \citenamefont
  {Centurion}}]{Yang:SD1:044101}%
  \BibitemOpen
  \bibfield  {author} {\bibinfo {author} {\bibfnamefont {J.}~\bibnamefont
  {Yang}}, \bibinfo {author} {\bibfnamefont {V.}~\bibnamefont {Makhija}},
  \bibinfo {author} {\bibfnamefont {V.}~\bibnamefont {Kumarappan}}, \ and\
  \bibinfo {author} {\bibfnamefont {M.}~\bibnamefont {Centurion}},\ }\href
  {\doibase 10.1063/1.4889840} {\bibfield  {journal} {\bibinfo  {journal}
  {Struct.\ Dyn.}\ }\textbf {\bibinfo {volume} {1}},\ \bibinfo {pages} {044101}
  (\bibinfo {year} {2014})}\BibitemShut {NoStop}%
\bibitem [{\citenamefont {Stapelfeldt}\ and\ \citenamefont
  {Seideman}(2003)}]{Stapelfeldt:RMP75:543}%
  \BibitemOpen
  \bibfield  {author} {\bibinfo {author} {\bibfnamefont {H.}~\bibnamefont
  {Stapelfeldt}}\ and\ \bibinfo {author} {\bibfnamefont {T.}~\bibnamefont
  {Seideman}},\ }\href {\doibase 10.1103/RevModPhys.75.543} {\bibfield
  {journal} {\bibinfo  {journal} {Rev.\ Mod.\ Phys.}\ }\textbf {\bibinfo
  {volume} {75}},\ \bibinfo {pages} {543} (\bibinfo {year} {2003})}\BibitemShut
  {NoStop}%
\bibitem [{\citenamefont {Holmegaard}\ \emph {et~al.}(2009)\citenamefont
  {Holmegaard}, \citenamefont {Nielsen}, \citenamefont {Nevo}, \citenamefont
  {Stapelfeldt}, \citenamefont {Filsinger}, \citenamefont {K{\"u}pper},\ and\
  \citenamefont {Meijer}}]{Holmegaard:PRL102:023001}%
  \BibitemOpen
  \bibfield  {author} {\bibinfo {author} {\bibfnamefont {L.}~\bibnamefont
  {Holmegaard}}, \bibinfo {author} {\bibfnamefont {J.~H.}\ \bibnamefont
  {Nielsen}}, \bibinfo {author} {\bibfnamefont {I.}~\bibnamefont {Nevo}},
  \bibinfo {author} {\bibfnamefont {H.}~\bibnamefont {Stapelfeldt}}, \bibinfo
  {author} {\bibfnamefont {F.}~\bibnamefont {Filsinger}}, \bibinfo {author}
  {\bibfnamefont {J.}~\bibnamefont {K{\"u}pper}}, \ and\ \bibinfo {author}
  {\bibfnamefont {G.}~\bibnamefont {Meijer}},\ }\href {\doibase
  10.1103/PhysRevLett.102.023001} {\bibfield  {journal} {\bibinfo  {journal}
  {Phys.\ Rev.\ Lett.}\ }\textbf {\bibinfo {volume} {102}},\ \bibinfo {pages}
  {023001} (\bibinfo {year} {2009})},\ \Eprint {http://arxiv.org/abs/0810.2307}
  {arXiv:0810.2307 [physics.chem-ph]} \BibitemShut {NoStop}%
\bibitem [{\citenamefont {Ghafur}\ \emph {et~al.}(2009)\citenamefont {Ghafur},
  \citenamefont {Rouzee}, \citenamefont {Gijsbertsen}, \citenamefont {Siu},
  \citenamefont {Stolte},\ and\ \citenamefont
  {Vrakking}}]{Ghafur:NatPhys5:289}%
  \BibitemOpen
  \bibfield  {author} {\bibinfo {author} {\bibfnamefont {O.}~\bibnamefont
  {Ghafur}}, \bibinfo {author} {\bibfnamefont {A.}~\bibnamefont {Rouzee}},
  \bibinfo {author} {\bibfnamefont {A.}~\bibnamefont {Gijsbertsen}}, \bibinfo
  {author} {\bibfnamefont {W.~K.}\ \bibnamefont {Siu}}, \bibinfo {author}
  {\bibfnamefont {S.}~\bibnamefont {Stolte}}, \ and\ \bibinfo {author}
  {\bibfnamefont {M.~J.~J.}\ \bibnamefont {Vrakking}},\ }\href {\doibase
  10.1038/nphys1225} {\bibfield  {journal} {\bibinfo  {journal} {Nat. Phys.}\
  }\textbf {\bibinfo {volume} {5}},\ \bibinfo {pages} {289} (\bibinfo {year}
  {2009})}\BibitemShut {NoStop}%
\bibitem [{\citenamefont {Kraus}\ \emph {et~al.}(2014)\citenamefont {Kraus},
  \citenamefont {Baykusheva},\ and\ \citenamefont
  {W{\"o}rner}}]{Kraus:PRL113:023001}%
  \BibitemOpen
  \bibfield  {author} {\bibinfo {author} {\bibfnamefont {P.~M.}\ \bibnamefont
  {Kraus}}, \bibinfo {author} {\bibfnamefont {D.}~\bibnamefont {Baykusheva}}, \
  and\ \bibinfo {author} {\bibfnamefont {H.~J.}\ \bibnamefont {W{\"o}rner}},\
  }\href {\doibase 10.1103/PhysRevLett.113.023001} {\bibfield  {journal}
  {\bibinfo  {journal} {Phys.\ Rev.\ Lett.}\ }\textbf {\bibinfo {volume}
  {113}},\ \bibinfo {pages} {023001} (\bibinfo {year} {2014})},\ \Eprint
  {http://arxiv.org/abs/1311.3923} {arXiv:1311.3923 [physics.chem-ph]}
  \BibitemShut {NoStop}%
\bibitem [{\citenamefont {Trippel}\ \emph {et~al.}(2015)\citenamefont
  {Trippel}, \citenamefont {Mullins}, \citenamefont {M{\"u}ller}, \citenamefont
  {Kienitz}, \citenamefont {Gonz{\'a}lez-F{\'e}rez},\ and\ \citenamefont
  {K{\"u}pper}}]{Trippel:PRL114:103003}%
  \BibitemOpen
  \bibfield  {author} {\bibinfo {author} {\bibfnamefont {S.}~\bibnamefont
  {Trippel}}, \bibinfo {author} {\bibfnamefont {T.}~\bibnamefont {Mullins}},
  \bibinfo {author} {\bibfnamefont {N.~L.~M.}\ \bibnamefont {M{\"u}ller}},
  \bibinfo {author} {\bibfnamefont {J.~S.}\ \bibnamefont {Kienitz}}, \bibinfo
  {author} {\bibfnamefont {R.}~\bibnamefont {Gonz{\'a}lez-F{\'e}rez}}, \ and\
  \bibinfo {author} {\bibfnamefont {J.}~\bibnamefont {K{\"u}pper}},\ }\href
  {\doibase 10.1103/PhysRevLett.114.103003} {\bibfield  {journal} {\bibinfo
  {journal} {Phys.\ Rev.\ Lett.}\ }\textbf {\bibinfo {volume} {114}},\ \bibinfo
  {pages} {103003} (\bibinfo {year} {2015})},\ \Eprint
  {http://arxiv.org/abs/1409.2836} {arXiv:1409.2836 [quant-ph]} \BibitemShut
  {NoStop}%
\bibitem [{\citenamefont {van Oudheusden}\ \emph {et~al.}(2010)\citenamefont
  {van Oudheusden}, \citenamefont {Pasmans}, \citenamefont {van~der Geer},
  \citenamefont {de~Loos}, \citenamefont {van~der Wiel},\ and\ \citenamefont
  {Luiten}}]{vanOudheusden:PRL105:264801}%
  \BibitemOpen
  \bibfield  {author} {\bibinfo {author} {\bibfnamefont {T.}~\bibnamefont {van
  Oudheusden}}, \bibinfo {author} {\bibfnamefont {P.~L. E.~M.}\ \bibnamefont
  {Pasmans}}, \bibinfo {author} {\bibfnamefont {S.~B.}\ \bibnamefont {van~der
  Geer}}, \bibinfo {author} {\bibfnamefont {M.~J.}\ \bibnamefont {de~Loos}},
  \bibinfo {author} {\bibfnamefont {M.~J.}\ \bibnamefont {van~der Wiel}}, \
  and\ \bibinfo {author} {\bibfnamefont {O.~J.}\ \bibnamefont {Luiten}},\
  }\href {\doibase 10.1103/PhysRevLett.105.264801} {\bibfield  {journal}
  {\bibinfo  {journal} {Phys.\ Rev.\ Lett.}\ }\textbf {\bibinfo {volume}
  {105}},\ \bibinfo {pages} {264801} (\bibinfo {year} {2010})}\BibitemShut
  {NoStop}%
\bibitem [{\citenamefont {Sciaini}\ and\ \citenamefont
  {Miller}(2011)}]{Sciaini:RPP74:096101}%
  \BibitemOpen
  \bibfield  {author} {\bibinfo {author} {\bibfnamefont {G.}~\bibnamefont
  {Sciaini}}\ and\ \bibinfo {author} {\bibfnamefont {R.~J.~D.}\ \bibnamefont
  {Miller}},\ }\href {\doibase 10.1088/0034-4885/74/9/096101} {\bibfield
  {journal} {\bibinfo  {journal} {Rep.\ Prog.\ Phys.}\ }\textbf {\bibinfo
  {volume} {74}},\ \bibinfo {pages} {096101} (\bibinfo {year}
  {2011})}\BibitemShut {NoStop}%
\bibitem [{\citenamefont {Musumeci}\ \emph {et~al.}(2010)\citenamefont
  {Musumeci}, \citenamefont {Moody}, \citenamefont {Scoby}, \citenamefont
  {Gutierrez}, \citenamefont {Bender},\ and\ \citenamefont
  {Wilcox}}]{Musumeci:RSI:013306}%
  \BibitemOpen
  \bibfield  {author} {\bibinfo {author} {\bibfnamefont {P.}~\bibnamefont
  {Musumeci}}, \bibinfo {author} {\bibfnamefont {J.~T.}\ \bibnamefont {Moody}},
  \bibinfo {author} {\bibfnamefont {C.~M.}\ \bibnamefont {Scoby}}, \bibinfo
  {author} {\bibfnamefont {M.~S.}\ \bibnamefont {Gutierrez}}, \bibinfo {author}
  {\bibfnamefont {H.~A.}\ \bibnamefont {Bender}}, \ and\ \bibinfo {author}
  {\bibfnamefont {N.~S.}\ \bibnamefont {Wilcox}},\ }\href {\doibase
  10.1063/1.3292683} {\bibfield  {journal} {\bibinfo  {journal} {Rev.\ Sci.\
  Instrum.}\ }\textbf {\bibinfo {volume} {81}},\ \bibinfo {pages} {013306}
  (\bibinfo {year} {2010})}\BibitemShut {NoStop}%
\bibitem [{\citenamefont {Robinson}\ \emph {et~al.}(2015)\citenamefont
  {Robinson}, \citenamefont {Lane},\ and\ \citenamefont
  {Wann}}]{Robinson:RSI86:013109}%
  \BibitemOpen
  \bibfield  {author} {\bibinfo {author} {\bibfnamefont {M.~S.}\ \bibnamefont
  {Robinson}}, \bibinfo {author} {\bibfnamefont {P.~D.}\ \bibnamefont {Lane}},
  \ and\ \bibinfo {author} {\bibfnamefont {D.~A.}\ \bibnamefont {Wann}},\
  }\href {\doibase 10.1063/1.4905335} {\bibfield  {journal} {\bibinfo
  {journal} {Rev.\ Sci.\ Instrum.}\ }\textbf {\bibinfo {volume} {86}},\
  \bibinfo {pages} {013109} (\bibinfo {year} {2015})}\BibitemShut {NoStop}%
\bibitem [{\citenamefont {Gerbig}\ \emph {et~al.}(2015)\citenamefont {Gerbig},
  \citenamefont {Senftleben}, \citenamefont {Morgenstern}, \citenamefont
  {Sarpe},\ and\ \citenamefont {Baumert}}]{Gerbig:NJP17:043050}%
  \BibitemOpen
  \bibfield  {author} {\bibinfo {author} {\bibfnamefont {C.}~\bibnamefont
  {Gerbig}}, \bibinfo {author} {\bibfnamefont {A.}~\bibnamefont {Senftleben}},
  \bibinfo {author} {\bibfnamefont {S.}~\bibnamefont {Morgenstern}}, \bibinfo
  {author} {\bibfnamefont {C.}~\bibnamefont {Sarpe}}, \ and\ \bibinfo {author}
  {\bibfnamefont {T.}~\bibnamefont {Baumert}},\ }\href {\doibase
  10.1088/1367-2630/17/4/043050} {\bibfield  {journal} {\bibinfo  {journal}
  {New J.\ Phys.}\ }\textbf {\bibinfo {volume} {17}},\ \bibinfo {pages}
  {043050} (\bibinfo {year} {2015})}\BibitemShut {NoStop}%
\bibitem [{\citenamefont {Lahme}\ \emph {et~al.}(2014)\citenamefont {Lahme},
  \citenamefont {Kealhofer}, \citenamefont {Krausz},\ and\ \citenamefont
  {Baum}}]{Lahme:SD1:034303}%
  \BibitemOpen
  \bibfield  {author} {\bibinfo {author} {\bibfnamefont {S.}~\bibnamefont
  {Lahme}}, \bibinfo {author} {\bibfnamefont {C.}~\bibnamefont {Kealhofer}},
  \bibinfo {author} {\bibfnamefont {F.}~\bibnamefont {Krausz}}, \ and\ \bibinfo
  {author} {\bibfnamefont {P.}~\bibnamefont {Baum}},\ }\href {\doibase
  10.1063/1.4884937} {\bibfield  {journal} {\bibinfo  {journal} {Struct.\
  Dyn.}\ }\textbf {\bibinfo {volume} {1}},\ \bibinfo {pages} {034303} (\bibinfo
  {year} {2014})}\BibitemShut {NoStop}%
\bibitem [{\citenamefont {van Mourik}\ \emph {et~al.}(2014)\citenamefont {van
  Mourik}, \citenamefont {Engelen}, \citenamefont {Vredenbregt},\ and\
  \citenamefont {Luiten}}]{vanMourik:SD1:034302}%
  \BibitemOpen
  \bibfield  {author} {\bibinfo {author} {\bibfnamefont {M.~W.}\ \bibnamefont
  {van Mourik}}, \bibinfo {author} {\bibfnamefont {W.~J.}\ \bibnamefont
  {Engelen}}, \bibinfo {author} {\bibfnamefont {E.~J.~D.}\ \bibnamefont
  {Vredenbregt}}, \ and\ \bibinfo {author} {\bibfnamefont {O.~J.}\ \bibnamefont
  {Luiten}},\ }\href {\doibase 10.1063/1.4882074} {\bibfield  {journal}
  {\bibinfo  {journal} {Struct.\ Dyn.}\ }\textbf {\bibinfo {volume} {1}},\
  \bibinfo {pages} {034302} (\bibinfo {year} {2014})}\BibitemShut {NoStop}%
\bibitem [{\citenamefont {McCulloch}\ \emph {et~al.}(2011)\citenamefont
  {McCulloch}, \citenamefont {Sheludko}, \citenamefont {Saliba}, \citenamefont
  {Bell}, \citenamefont {Junker}, \citenamefont {Nugent},\ and\ \citenamefont
  {Scholten}}]{McCulloch:NatPhys10:785}%
  \BibitemOpen
  \bibfield  {author} {\bibinfo {author} {\bibfnamefont {A.~J.}\ \bibnamefont
  {McCulloch}}, \bibinfo {author} {\bibfnamefont {D.~V.}\ \bibnamefont
  {Sheludko}}, \bibinfo {author} {\bibfnamefont {S.~D.}\ \bibnamefont
  {Saliba}}, \bibinfo {author} {\bibfnamefont {S.~C.}\ \bibnamefont {Bell}},
  \bibinfo {author} {\bibfnamefont {M.}~\bibnamefont {Junker}}, \bibinfo
  {author} {\bibfnamefont {K.~A.}\ \bibnamefont {Nugent}}, \ and\ \bibinfo
  {author} {\bibfnamefont {R.~E.}\ \bibnamefont {Scholten}},\ }\href {\doibase
  10.1038/nphys2052} {\bibfield  {journal} {\bibinfo  {journal} {Nat. Phys.}\
  }\textbf {\bibinfo {volume} {7}},\ \bibinfo {pages} {785} (\bibinfo {year}
  {2011})}\BibitemShut {NoStop}%
\bibitem [{\citenamefont {Gulde}\ \emph {et~al.}(2014)\citenamefont {Gulde},
  \citenamefont {Schweda}, \citenamefont {Storeck}, \citenamefont {Maiti},
  \citenamefont {Yu}, \citenamefont {Wodtke}, \citenamefont {Sch{\"a}fer},\
  and\ \citenamefont {Ropers}}]{Gulde:Science345:200}%
  \BibitemOpen
  \bibfield  {author} {\bibinfo {author} {\bibfnamefont {M.}~\bibnamefont
  {Gulde}}, \bibinfo {author} {\bibfnamefont {S.}~\bibnamefont {Schweda}},
  \bibinfo {author} {\bibfnamefont {G.}~\bibnamefont {Storeck}}, \bibinfo
  {author} {\bibfnamefont {M.}~\bibnamefont {Maiti}}, \bibinfo {author}
  {\bibfnamefont {H.~K.}\ \bibnamefont {Yu}}, \bibinfo {author} {\bibfnamefont
  {A.~M.}\ \bibnamefont {Wodtke}}, \bibinfo {author} {\bibfnamefont
  {S.}~\bibnamefont {Sch{\"a}fer}}, \ and\ \bibinfo {author} {\bibfnamefont
  {C.}~\bibnamefont {Ropers}},\ }\href {\doibase 10.1126/science.1250658}
  {\bibfield  {journal} {\bibinfo  {journal} {Science}\ }\textbf {\bibinfo
  {volume} {345}},\ \bibinfo {pages} {200} (\bibinfo {year}
  {2014})}\BibitemShut {NoStop}%
\bibitem [{\citenamefont {Zuo}\ \emph {et~al.}(1996)\citenamefont {Zuo},
  \citenamefont {Bandrauk},\ and\ \citenamefont {Corkum}}]{Zuo:CPL159:313}%
  \BibitemOpen
  \bibfield  {author} {\bibinfo {author} {\bibfnamefont {T.}~\bibnamefont
  {Zuo}}, \bibinfo {author} {\bibfnamefont {A.~D.}\ \bibnamefont {Bandrauk}}, \
  and\ \bibinfo {author} {\bibfnamefont {P.~B.}\ \bibnamefont {Corkum}},\
  }\href@noop {} {\bibfield  {journal} {\bibinfo  {journal} {Chem.\ Phys.\
  Lett.}\ }\textbf {\bibinfo {volume} {259}},\ \bibinfo {pages} {313} (\bibinfo
  {year} {1996})}\BibitemShut {NoStop}%
\bibitem [{\citenamefont {Blaga}\ \emph {et~al.}(2012)\citenamefont {Blaga},
  \citenamefont {Xu}, \citenamefont {DiChiara}, \citenamefont {Sistrunk},
  \citenamefont {Zhang}, \citenamefont {Agostini}, \citenamefont {Miller},
  \citenamefont {DiMauro},\ and\ \citenamefont {Lin}}]{Blaga:Nature483:194}%
  \BibitemOpen
  \bibfield  {author} {\bibinfo {author} {\bibfnamefont {C.~I.}\ \bibnamefont
  {Blaga}}, \bibinfo {author} {\bibfnamefont {J.}~\bibnamefont {Xu}}, \bibinfo
  {author} {\bibfnamefont {A.~D.}\ \bibnamefont {DiChiara}}, \bibinfo {author}
  {\bibfnamefont {E.}~\bibnamefont {Sistrunk}}, \bibinfo {author}
  {\bibfnamefont {K.}~\bibnamefont {Zhang}}, \bibinfo {author} {\bibfnamefont
  {P.}~\bibnamefont {Agostini}}, \bibinfo {author} {\bibfnamefont {T.~A.}\
  \bibnamefont {Miller}}, \bibinfo {author} {\bibfnamefont {L.~F.}\
  \bibnamefont {DiMauro}}, \ and\ \bibinfo {author} {\bibfnamefont {C.~D.}\
  \bibnamefont {Lin}},\ }\href@noop {} {\bibfield  {journal} {\bibinfo
  {journal} {Nature}\ }\textbf {\bibinfo {volume} {483}},\ \bibinfo {pages}
  {194} (\bibinfo {year} {2012})}\BibitemShut {NoStop}%
\bibitem [{\citenamefont {Chang}\ \emph {et~al.}(2013)\citenamefont {Chang},
  \citenamefont {D{\l}ugo\l\k{e}cki}, \citenamefont {K{\"u}pper}, \citenamefont
  {R{\"o}sch}, \citenamefont {Wild},\ and\ \citenamefont
  {Willitsch}}]{Chang:Science342:98}%
  \BibitemOpen
  \bibfield  {author} {\bibinfo {author} {\bibfnamefont {Y.-P.}\ \bibnamefont
  {Chang}}, \bibinfo {author} {\bibfnamefont {K.}~\bibnamefont
  {D{\l}ugo\l\k{e}cki}}, \bibinfo {author} {\bibfnamefont {J.}~\bibnamefont
  {K{\"u}pper}}, \bibinfo {author} {\bibfnamefont {D.}~\bibnamefont
  {R{\"o}sch}}, \bibinfo {author} {\bibfnamefont {D.}~\bibnamefont {Wild}}, \
  and\ \bibinfo {author} {\bibfnamefont {S.}~\bibnamefont {Willitsch}},\ }\href
  {\doibase 10.1126/science.1242271} {\bibfield  {journal} {\bibinfo  {journal}
  {Science}\ }\textbf {\bibinfo {volume} {342}},\ \bibinfo {pages} {98}
  (\bibinfo {year} {2013})},\ \Eprint {http://arxiv.org/abs/1308.6538}
  {arXiv:1308.6538 [physics]} \BibitemShut {NoStop}%
\bibitem [{\citenamefont {Trippel}\ \emph {et~al.}(2013)\citenamefont
  {Trippel}, \citenamefont {Mullins}, \citenamefont {M{\"u}ller}, \citenamefont
  {Kienitz}, \citenamefont {D{\l}ugo{\l}\k{e}cki},\ and\ \citenamefont
  {K{\"u}pper}}]{Trippel:MP111:1738}%
  \BibitemOpen
  \bibfield  {author} {\bibinfo {author} {\bibfnamefont {S.}~\bibnamefont
  {Trippel}}, \bibinfo {author} {\bibfnamefont {T.}~\bibnamefont {Mullins}},
  \bibinfo {author} {\bibfnamefont {N.~L.~M.}\ \bibnamefont {M{\"u}ller}},
  \bibinfo {author} {\bibfnamefont {J.~S.}\ \bibnamefont {Kienitz}}, \bibinfo
  {author} {\bibfnamefont {K.}~\bibnamefont {D{\l}ugo{\l}\k{e}cki}}, \ and\
  \bibinfo {author} {\bibfnamefont {J.}~\bibnamefont {K{\"u}pper}},\ }\href
  {\doibase 10.1080/00268976.2013.780334} {\bibfield  {journal} {\bibinfo
  {journal} {Mol.\ Phys.}\ }\textbf {\bibinfo {volume} {111}},\ \bibinfo
  {pages} {1738} (\bibinfo {year} {2013})},\ \Eprint
  {http://arxiv.org/abs/1301.1826} {arXiv:1301.1826 [physics.atom-ph]}
  \BibitemShut {NoStop}%
\bibitem [{\citenamefont {Eppink}\ and\ \citenamefont
  {Parker}(1997)}]{Eppink:RSI68:3477}%
  \BibitemOpen
  \bibfield  {author} {\bibinfo {author} {\bibfnamefont {A.~T. J.~B.}\
  \bibnamefont {Eppink}}\ and\ \bibinfo {author} {\bibfnamefont {D.~H.}\
  \bibnamefont {Parker}},\ }\href {\doibase 10.1063/1.1148310} {\bibfield
  {journal} {\bibinfo  {journal} {Rev.\ Sci.\ Instrum.}\ }\textbf {\bibinfo
  {volume} {68}},\ \bibinfo {pages} {3477} (\bibinfo {year}
  {1997})}\BibitemShut {NoStop}%
\bibitem [{\citenamefont {Stei}\ \emph {et~al.}(2013)\citenamefont {Stei},
  \citenamefont {von Vangerow}, \citenamefont {Otto}, \citenamefont {Kelkar},
  \citenamefont {Carrascosa}, \citenamefont {Best},\ and\ \citenamefont
  {Wester}}]{Stei:JCP138:214201}%
  \BibitemOpen
  \bibfield  {author} {\bibinfo {author} {\bibfnamefont {M.}~\bibnamefont
  {Stei}}, \bibinfo {author} {\bibfnamefont {J.}~\bibnamefont {von Vangerow}},
  \bibinfo {author} {\bibfnamefont {R.}~\bibnamefont {Otto}}, \bibinfo {author}
  {\bibfnamefont {A.~H.}\ \bibnamefont {Kelkar}}, \bibinfo {author}
  {\bibfnamefont {E.}~\bibnamefont {Carrascosa}}, \bibinfo {author}
  {\bibfnamefont {T.}~\bibnamefont {Best}}, \ and\ \bibinfo {author}
  {\bibfnamefont {R.}~\bibnamefont {Wester}},\ }\href@noop {} {\bibfield
  {journal} {\bibinfo  {journal} {J.\ Chem.\ Phys.}\ }\textbf {\bibinfo
  {volume} {138}},\ \bibinfo {pages} {214201} (\bibinfo {year}
  {2013})}\BibitemShut {NoStop}%
\bibitem [{\citenamefont {Bainbridge}\ and\ \citenamefont
  {Bryan}(2014)}]{Bainbridge:NJP16:103031}%
  \BibitemOpen
  \bibfield  {author} {\bibinfo {author} {\bibfnamefont {A.~R.}\ \bibnamefont
  {Bainbridge}}\ and\ \bibinfo {author} {\bibfnamefont {W.~A.}\ \bibnamefont
  {Bryan}},\ }\href {\doibase 10.1088/1367-2630/16/10/103031} {\bibfield
  {journal} {\bibinfo  {journal} {New J.\ Phys.}\ }\textbf {\bibinfo {volume}
  {16}},\ \bibinfo {pages} {103031} (\bibinfo {year} {2014})}\BibitemShut
  {NoStop}%
\bibitem [{\citenamefont {Maldonado}\ \emph {et~al.}(2012)\citenamefont
  {Maldonado}, \citenamefont {Pianetta}, \citenamefont {Dowell}, \citenamefont
  {Corbett}, \citenamefont {Park}, \citenamefont {Schmerge}, \citenamefont
  {Trautwein},\ and\ \citenamefont {Clay}}]{Maldonado:APL101:231103}%
  \BibitemOpen
  \bibfield  {author} {\bibinfo {author} {\bibfnamefont {J.~R.}\ \bibnamefont
  {Maldonado}}, \bibinfo {author} {\bibfnamefont {P.}~\bibnamefont {Pianetta}},
  \bibinfo {author} {\bibfnamefont {D.~H.}\ \bibnamefont {Dowell}}, \bibinfo
  {author} {\bibfnamefont {J.}~\bibnamefont {Corbett}}, \bibinfo {author}
  {\bibfnamefont {S.}~\bibnamefont {Park}}, \bibinfo {author} {\bibfnamefont
  {J.}~\bibnamefont {Schmerge}}, \bibinfo {author} {\bibfnamefont
  {A.}~\bibnamefont {Trautwein}}, \ and\ \bibinfo {author} {\bibfnamefont
  {W.}~\bibnamefont {Clay}},\ }\href {\doibase 10.1063/1.4769220} {\bibfield
  {journal} {\bibinfo  {journal} {Astrophys.\ Lett.\ \& Comm.}\ }\textbf
  {\bibinfo {volume} {101}},\ \bibinfo {pages} {231103} (\bibinfo {year}
  {2012})}\BibitemShut {NoStop}%
\bibitem [{\citenamefont {Floettmann}(1997)}]{Floettmann:ASTRA:1997}%
  \BibitemOpen
  \bibfield  {author} {\bibinfo {author} {\bibfnamefont {K.}~\bibnamefont
  {Floettmann}},\ }\href {http://www.desy.de/~mpyflo/} {\enquote {\bibinfo
  {title} {{ASTRA}, a space charge tracking algorithm},}\ } (\bibinfo {year}
  {1997})\BibitemShut {NoStop}%
\bibitem [{\citenamefont {Fl\"ugge}(1986)}]{Fluegge:Elektrodynamik:1986}%
  \BibitemOpen
  \bibfield  {author} {\bibinfo {author} {\bibfnamefont {S.}~\bibnamefont
  {Fl\"ugge}},\ }\href {\doibase 10.1007/978-3-642-71143-5} {\emph {\bibinfo
  {title} {Rechenmethoden der Elektrodynamik}}},\ Aufgaben mit L\"osungen\
  (\bibinfo  {publisher} {Springer Verlag},\ \bibinfo {year}
  {1986})\BibitemShut {NoStop}%
\bibitem [{\citenamefont {Trippel}\ \emph {et~al.}(2014)\citenamefont
  {Trippel}, \citenamefont {Mullins}, \citenamefont {M{\"u}ller}, \citenamefont
  {Kienitz}, \citenamefont {Omiste}, \citenamefont {Stapelfeldt}, \citenamefont
  {Gonz{\'a}lez-F{\'e}rez},\ and\ \citenamefont
  {K{\"u}pper}}]{Trippel:PRA89:051401R}%
  \BibitemOpen
  \bibfield  {author} {\bibinfo {author} {\bibfnamefont {S.}~\bibnamefont
  {Trippel}}, \bibinfo {author} {\bibfnamefont {T.}~\bibnamefont {Mullins}},
  \bibinfo {author} {\bibfnamefont {N.~L.~M.}\ \bibnamefont {M{\"u}ller}},
  \bibinfo {author} {\bibfnamefont {J.~S.}\ \bibnamefont {Kienitz}}, \bibinfo
  {author} {\bibfnamefont {J.~J.}\ \bibnamefont {Omiste}}, \bibinfo {author}
  {\bibfnamefont {H.}~\bibnamefont {Stapelfeldt}}, \bibinfo {author}
  {\bibfnamefont {R.}~\bibnamefont {Gonz{\'a}lez-F{\'e}rez}}, \ and\ \bibinfo
  {author} {\bibfnamefont {J.}~\bibnamefont {K{\"u}pper}},\ }\href {\doibase
  10.1103/PhysRevA.89.051401} {\bibfield  {journal} {\bibinfo  {journal}
  {Phys.\ Rev.\ A}\ }\textbf {\bibinfo {volume} {89}},\ \bibinfo {pages}
  {051401(R)} (\bibinfo {year} {2014})},\ \Eprint
  {http://arxiv.org/abs/1401.6897} {arXiv:1401.6897 [quant-ph]} \BibitemShut
  {NoStop}%
\bibitem [{\citenamefont {Fu}\ \emph {et~al.}(2014)\citenamefont {Fu},
  \citenamefont {Liu}, \citenamefont {Zhu}, \citenamefont {Xiang},
  \citenamefont {Zhang},\ and\ \citenamefont {Cao}}]{Fu:RIS85:083701}%
  \BibitemOpen
  \bibfield  {author} {\bibinfo {author} {\bibfnamefont {F.}~\bibnamefont
  {Fu}}, \bibinfo {author} {\bibfnamefont {S.}~\bibnamefont {Liu}}, \bibinfo
  {author} {\bibfnamefont {P.}~\bibnamefont {Zhu}}, \bibinfo {author}
  {\bibfnamefont {D.}~\bibnamefont {Xiang}}, \bibinfo {author} {\bibfnamefont
  {J.}~\bibnamefont {Zhang}}, \ and\ \bibinfo {author} {\bibfnamefont
  {J.}~\bibnamefont {Cao}},\ }\href {\doibase 10.1063/1.4892135} {\bibfield
  {journal} {\bibinfo  {journal} {Rev.\ Sci.\ Instrum.}\ }\textbf {\bibinfo
  {volume} {85}},\ \bibinfo {pages} {083701} (\bibinfo {year}
  {2014})}\BibitemShut {NoStop}%
\end{thebibliography}%

\end{document}